\documentclass[final,5p,times,twocolumn]{elsarticle}

\usepackage{amssymb}
\usepackage{amsmath}
\usepackage{url}
\usepackage{framed}
\usepackage{booktabs}
\usepackage{longtable}

\usepackage{xcolor}
\usepackage{booktabs}
\usepackage{tabularx}
\usepackage{rotating}

\journal{Journal of Systems and Software}

\begin{document}

\begin{frontmatter}


\title{Public Sector Open Source Program Offices - Archetypes for how to Grow (Common) Institutional Capabilities}


\author[1]{Johan Linåker}\corref{cor1}
\ead{johan.linaker@ri.se}
\ead[url]{https://orcid.org/0000-0001-9851-1404}

\author[2]{Astor Nummelin Carlberg}
\ead{astor@openforumeurope.org}

\author[2]{Ciarán O’Riordan}
\ead{ciaran @openforumeurope.org}

\cortext[cor1]{Corresponding author.}

\address[1]{RISE Research Institutes of Sweden AB, Lund, Sweden}
\address[2]{OpenForum Europe, Brussels, Belgium}




\begin{abstract}
\textbf{Context:} Open Source Software (OSS) is a crucial component of over 90\% of digital infrastructure underpinning industry and public digital services, facilitating collaborative software development and dissemination. Its significance in the European public sector has been emphasised through various Ministerial Declarations, highlighting its potential to accelerate digitalisation, transform businesses, and foster a digitally skilled population.
\textbf{Research Aim:}
This study aims to explore how the adoption, development, and collaboration on OSS can be enabled through organisational support functions or centres of competency, also known as Open Source Programme Offices (OSPOs) within Public Sector Organisations (PSOs) in the European Union, Norway, Liechtenstein, and Iceland.
\textbf{Methodology:}
A qualitative research approach was adopted, involving an interview survey of 18 OSPO representatives across 16 cases of public-sector OSPOs. These cases were cross-analysed and categorised into six OSPO archetypes. The findings were validated and enriched through two follow-up focus groups that included earlier interviewees and additional experts.
\textbf{Results:}
The study identified six distinct OSPO archetypes, providing insights into their organisational structures, responsibilities, and contributions to OSS adoption. The archetypes, along with policy recommendations, offer guidance on how PSOs can design their own OSPOs, taking into account their specific context, resources, and policy goals.
\textbf{Conclusions:}
The findings enhance the understanding of OSPOs as strategic endeavours aimed at promoting OSS adoption. The study offers practical guidance for PSOs and policymakers on leveraging OSS to achieve strategic objectives, foster digital sovereignty, drive economic growth, and improve the interoperability and quality of digital services.
\end{abstract}




\end{frontmatter}




\section{Introduction}
\label{sec:introduction}

Open Source Software (OSS) provides a mechanism for collaborative software development and dissemination, as well as a type of building block present in over 90 percent of digital infrastructure underpinning industry and public digital services~\cite{synopsis2025survey, github2024octoverse}. Its importance in the European public sector domain has been stressed through several Ministerial Declarations~\cite{ec2017tallin, ec2020berlin, ec2022strasbourg}, highlighting the potential of OSS to accelerate the digitalisation of public services, digital transformation of businesses, and the growth of a digitally skilled workforce as defined in the European Digital Compass~\cite{ec2021digitalcompass}. These arguments are supported by studies, for example, finding that OSS uptake and competence are essential for a region's or nation's digital autonomy and for economic growth in terms of start-ups, jobs, and GDP~\cite{blind2021impact, nagle2019government}. For digital services and products specifically, OSS is a tool to improve interoperability and quality while increasing transparency and accountability~\cite{ofe2022strategic}.

Public Sector Organisations (PSOs) are, however, commonly faced with numerous challenges in terms of leveraging the potential of OSS as an instrument in their digital transformations~\cite{linaaker2025public}. One such challenge regards their reliance on the acquisition and outsourcing of technical capabilities~\cite{marco2020outsourcing}, which has limited the development of competencies regarding OSS~\cite{linaaker2020public} and software engineering practices~\cite{borg2018digitalization}. Regulatory constraints and complicated procurement frameworks~\cite{lundell2021enabling}, along with limited or short-term policy incentives~\cite{alanne2015three}, and a conservative IT sourcing culture~\cite{valimaki2005empirical} further complicate matters. These challenges are particularly pronounced for municipalities due to their smaller size and limited resources, highlighting their need for support~\cite{shaikh2016negotiating, van2015adopting, mergel2016agile}.

Within industry, an established approach to addressing these challenges has been to establish centres of competency, commonly referred to as Open Source Programme Offices (OSPOs)~\cite{todo2024definition}. Their structure and responsibilities vary depending on an organisation's needs, goals, and conditions~\cite{haddad2020ospos}. However, they typically share the common goal of enabling the use of OSS on the strategic and operational levels in line with the overarching business goals of an organisation~\cite{munir2021rise}.

Since the OSPOs first appeared in the 2000s~\cite{ruff2022rise, phipps2021sunsospo, capek2005history}, their adoption has steadily increased. Today, more than 30\% of Fortune 100 companies have established an OSPO~\cite{github2024octoverse}, and the adoption is generally trending upwards~\cite{todogroup2024survey}. In the public sector, the phenomenon of establishing OSPOs has recently emerged. For example, the recent US Securing Open Source Act recommends that US federal agencies set up OSPOs~\cite{uscongress2022securing}.

In Europe, there are no formal recommendations or requirements, although there are some exceptions~\cite{ec2020ossStrategy, bothorell2021mission}. Yet, there are several examples of OSPOs being reported both on supranational (for example, European Commission\footnote{\url{https://interoperable-europe.ec.europa.eu/collection/ec-ospo}}), national (for example, Italy\footnote{\url{https://developers.italia.it/}}), regional (for example, Galicia\footnote{\url{https://mancomun.gal/es/}}), and local (for example, Paris\footnote{\url{https://opensource.paris.fr/ossparis/english.html}}) levels of government, as well as in state-owned enterprises (for example, Alliander~\footnote{\url{https://www.alliander.com/en/open-source/}}), academia (for example, Trinity College Dublin\footnote{\url{https://www.tcd.ie/innovation/for-trinity-innovators/open-source-programme-office/}}), and international humanitarian organisations (for example, WHO~\footnote{\url{https://pandemichub.who.int/}}). The latter is recognised specifically in the context of how OSPOs can help tackle humanitarian and social challenges through the UN's yearly OSPOs for Good conference~\footnote{\url{https://www.un.org/digital-emerging-technologies/content/ospos-good-2024}}.

The OSPO label is not used consistently across the examples, yet they all share similar features, which can be compared to many definitions of the OSPO construct~\cite{todo2024definition}. Research on how OSPOs are structured and organised, and what responsibilities they take on, is limited to a few notable examples. 

Herpig~\cite{herpig2023ospos}) explores how Cyber Security OSPOs can be set up as a part of an overarching public sector OSPO or an independent unit inside a national cybersecurity agency. The focus of such an OSPO includes “supply-chain security policy work, [and] coordinating across government agencies and offices to ensure that OSS security is no longer tied to a crisis cycle”. OpenForum Europe~\cite{openforumeurope2022ospo} explores four cases of public sector OSPOs mainly from a policy perspective, highlighting underpinning goals and strategies of the hosting PSOs. The Linux Foundation~\cite{linuxfoundation2024survey} surveys the state of OSS in the European public sector, highlighting OSPOs as a key enabler for OSS operations and adoption, drawing on insights from several OSPO interviewees. None of the reports, however, provides guidance for the public sector at large in terms of how OSPOs may be designed, considering contextual factors, such as the level of government, resources, and policy goals. 

Existing guidelines and resources are practically oriented, providing guidance on governance, processes, and tasks a general OSPO or organisation may take on when considering OSS as a strategic initiative~\cite{todo2024definition, ospoalliance2022goodgovernance, foundationforpubliccode2025standard}.

Academic literature has recognised their importance only to a limited extent~\cite{harutyunyan2019industry} and explored their representation in the industry context~\cite{munir2021rise, xia2023lessons, riehle2024open}. Digital government research has meanwhile highlighted the potential for open and agile collaboration in governments through social coding platforms~\cite{mergel2015open, mergel2016agile}, while also recognising challenges and the need for dedicated experts that can support OSS adoption inside PSOs (similar to the OSPO construct)~\cite{shaikh2016negotiating, van2015adopting}. Research has further focused on exploring the adoption of OSS solutions, for example, the organisational transition from proprietary to OSS alternatives~\cite{silic2017open, waring2005open}, and on identifying and evaluating potential benefits, risks, and challenges~\cite{koloniaris2018possibilities, deller2008open}. General factors highlighted include the clear need for technical~\cite{oram2011promoting} and procurement-related support~\cite{shaikh2016negotiating, mergel2016agile} and sustainable strategic commitment from policymakers~\cite{cassell2008governments, van2015adopting}. The latter can easily dissipate due to changes in public debate and government~\cite{maldonado2010process}.

This study addresses the gap in research and practice by exploring how the adoption, development, and collaboration on OSS can be enabled specifically through organisational support functions or centres of competency using the OSPO construct. The study is based on an underpinning report by the authors, commissioned by the 
Directorate-General for Digital Services (DIGIT) of the European Commission to 
OpenForum Europe and RISE through the OSOR project under Specific Contract 32 FWC DI/07929 with Wavestone SA. 

The study thereby has an explicit focus on how PSOs within the European Union (EU), Norway, Liechtenstein, and Iceland work to support and enable the consumption, development, and collaboration of OSS. A qualitative research approach is adopted, with an interview survey of 18 OSPO representatives across 16 cases of different public-sector OSPOs. These cases were cross-analysed and categorised into six OSPO archetypes, providing an overview of how the OSPOs within each archetype are organised, what their responsibilities entail, and how they support OSS within their respective scope and domain. Findings were validated and enriched through two follow-up focus groups that included earlier interviewees and additional experts. We make the following contributions:

\begin{itemize}
    \item An empirically grounded typology of OSPO archetypes and practice that can guide PSOs on the design of their own OSPOs, considering their own context, resources, and policy goals. 
    \item Recommendations for policy for the implementation of support structures and programs to leverage the potential OSS brings as a vehicle for digital innovation and sovereignty.
    \item A foundation and several directions for future digital government research on how the implementation of OSS policy can be supported in various contexts of the public sector.
\end{itemize}

As this is a qualitative study, readers should specifically consider the contextual descriptions of the reported cases before making any anecdotal generalisations. Further, the presented OSPO archetypes are based on a limited sample of studied cases, and provide only a partial picture of how public-sector OSPOs can be organised and structured. Each OSPO will be unique to its organisation's culture, focus, and practices. Still, the presented archetypes provide a source of inspiration and design knowledge for both practitioners and researchers.

The remainder of this report is structured as follows: first, background and related work are presented, followed by a conceptual framework based on this, which provides an a priori frame for the study (see Section 2). The research design is then introduced along with a discussion of the threats to the validity of the study (see Section 3). The results section is then provided, presenting the OSPO archetypes and the cross-analysis of cases studied (see Section 4), followed by the discussion and conclusions based on the findings (see Section 5).

\section{Background and related work}
Below, background and related work is presented that frame and provide input to the study, as well as the conceptual framework that is used as a foundation and framing for the continued investigation.

First, what constitutes an OSPO is defined based on a set of established sources. Then an overview is provided of how an OSPO may be structured, what roles may be included, and what responsibilities may lie with the OSPO in an industrial context. Following that, context is provided from an earlier report on public sector OSPOs from a public policy perspective. Finally, the conceptual framework is briefly presented.

\subsection{What defines an OSPO?}
The concept of OSPOs as a construct originates from industry at the beginning of the 2000s among larger software intensive companies such as Sun Microsystems, Hewlett-Packard, Intel, and Google~\cite{ruff2022rise, phipps2021sunsospo}. Since then, the concept has grown and matured as several organisations and industries, also outside of the software industry, are adopting the construct~\cite{todogroup2024survey}. 

With the growth of OSPOs, so has the definition and views of what an OSPO is. The TODO Group, an industry network, describes an OSPO as a \textit{``center of competency for an organisation's open source operations and structure''} which may include responsibilities such as designing and facilitating processes for selection and intake of OSS, compliance and auditing, contributions and community engagement, as well as training for an organisation's different stakeholders~\cite{todo2024definition}. The OSPO Alliance (a joint initiative by for example, OW2, Eclipse Foundation, OpenForum Europe, and Foundation for Public Code) considers an OSPO as \textit{``a cross-functional team to help define and steer an organisation's open source management strategy and organisational readiness''}~\cite{ospoalliance2021positionpaper}. Both networks agree that there is no ''one-size-fits-all''. Rather, its tasks and responsibilities will vary in line with the overarching organisation's goals, strategies, and views of OSS.

From a public sector perspective, the concept is still emerging, and commonly under different naming conventions, although with similar meanings. OpenForum Europe, for example, considers it as \textit{``an institutional organisational construct that supports and accelerates the consumption, creation, and application of Open Source software''}~\cite{openforumeurope2022ospo}. Specifically, they highlight the focus on \textit{``working strategically to achieve the policy objectives of the institution that intersect with Open Source''}, which contrasts with the business perspective in industry OSPOs. However, from a general perspective, the focus is still related to enabling OSS as a tool for creating value based on the goals of the focal organisation, public or private.

OSPO++, a network and community of collaborative OSPOs universities, governments, and civic institutions, accordingly looks at the construct from a public sector perspective, highlighting both municipal, government, and university settings. They consider how the public context can bring other organisational structures, incentives, and levels of bureaucracy than the private context~\cite{ospoplusplus2020ospos}. The common trend of outsourcing and acquiring technical capabilities~\cite{mikalsen2020can}, along with otherwise symptomatic short-term planning, and risk-aversive culture~\cite{alanne2015three}, further adds nuance that might affect how OSPOs in PSOs take shape. 

Taking these nuances into consideration, this study uses the OSPO-construct to study how PSOs support and enable the growth of their own capabilities related to consumption of, development of, and collaboration on OSS. Hence, the unit of analysis is the support functions that PSOs provide, not limited by how it is organised, or under what label it is referred to.

\subsection{Structures of an OSPO in the industry}
\label{subsec:OSPO-structures}
The structure and organisational location of an OSPO vary depending on its goals and purpose. Haddad~\cite{haddad2020ospos} reports on five examples of how an OSPO can be set up in an industrial setting (see Table~\ref{tbl:structure-of-ospos}). The examples vary in terms of their internal sponsor, budget, staffing, which entities they support, and the size of the organisation they are suitable for, as presented in the table below. 

\begin{table*}[t]
\caption{Overview of different OSPO structures observed in the industry setting. Adopted from Haddad~\cite{haddad2020ospos}.}
\label{tbl:structure-of-ospos}
\begin{tabular}{p{2.5cm}p{2.5cm}p{2.4cm}p{2.4cm}p{2.5cm}p{2.5cm}}
\toprule
 & \textbf{Sponsor} & \textbf{Supporting entities} & \textbf{Budget} & \textbf{Staffing} & \textbf{Org.} \\ \midrule
\textbf{OSPO within R\&D org.} & Executive mgmt for R\&D & R\&D division and product departments & Dedicated & Head of open source + dedicated staff & Large organisations with single product division \\ \midrule
\textbf{Corporate-level OSPO with division support} & Executive mgmt & Multiple divisions & Dedicated & Head of open source + dedicated staff & Large organisations with multiple product division \\ \midrule
\textbf{Virtual OSPO} & CTOs office or head of engineering & Main organisation & Partial funding across different functions & Head of OSS + part time distributed team & Small and medium-sized organisations \\ \midrule
\textbf{No official OSPO} & Executive mgmt & Main organisation & Partial funding across different functions & Part time distributed team & Small organisations \\ \midrule
\textbf{OSPO as part of CTO office or Engineering} & CTOs office or head of engineering & Main organisation & Dedicated & Head of open source + dedicated staff & Medium-sized organisations \\ 
\bottomrule
\end{tabular}
\end{table*}

The first example puts the OSPO within the R\&D organisation with direct sponsorship and reporting to executive management. This setup isolates the OSPO from product departments, enabling it to execute on a general OSS strategy not limited by specific product strategies. More freedom to collaborate with external parties is also given. 

The second yet similar example places the OSPO at the corporate level with support functions in each division when multiple product divisions are present. The support functions help adapt and execute OSS strategies, policies, and processes for the respective divisions' contexts. In a third example, a virtual OSPO, there is typically a main responsible head of OSS but without a fully dedicated team. Individuals from different functions instead work part-time on related tasks coordinated by the head of OSS.

In a fourth example, there is no official OSPO present. Rather, the associated tasks are spread out over multiple individuals, a suitable option for smaller organisations. In the fifth and final example, the OSPO is positioned as a part of the CTO office or the engineering organisation, suitable for medium-sized organisations~\cite{haddad2020ospos}. 

\subsection{Roles within an OSPO in the industry}
As with the structure and organisational location, the staffing of the OSPO depends on its goals and purpose. Haddad~\cite{haddad2020ospos} differentiates between five different roles, although highlighting that they do not necessarily need to be divided among different individuals. Rather, they can be either shared or aggregated, considering the structure of the OSPO. Roles include:

\begin{itemize}
 \item A head of OSS is responsible for the design and execution of the organisation's OSS strategy. The head also manages the reporting and follow-up of the strategy’s progress to the OSPO's sponsor. 
 \item A software architect is proposed with responsibility for high-level technical decisions related to OSS and the organisation's various products and platforms. 
 \item A technical evangelist is proposed to support and promote OSS contributions, engagement, and culture across the organisations and externally in relevant communities and foundations. 
 \item A compliance engineer is proposed to oversee and maintain compliance related to the organisation's OSS intake. 
 \item A legal counsel is proposed that is knowledgeable in OSS licensing.
\end{itemize}

In their classification, the TODO Group differentiates between roles focusing on governance, project management, licensing, security, community engagement, developer education, individual contributors, and OSS advice~\cite{todo2024definition}.

\subsection{Responsibilities of an OSPO in the industry}
\label{subsec:OSPO-responsibilities}
The TODO Group has defined a set of responsibilities typically maintained by OSPOs based on industry practice and an underpinning report by Haddad~\cite{haddad2020ospos}. Exactly which responsibilities an OSPO maintains depends on its purpose. Further, as responsibilities originate from state-of-practice in an industry context, not all may be considered relevant or correctly defined for the public sector context. Still, they provide a point of departure when investigating what responsibilities public sector OSPOs take on. Below, we briefly summarise describe an OSPO’s responsibilities primarily based on the work by Haddad~\cite{haddad2020ospos}:

\begin{itemize}

 \item \textbf{Develop and Execute OSS Strategy} - An OSPO is responsible for setting and implementing an organisation's long-term OSS strategy. This involves defining goals related to OSS consumption, participation, contribution, and leadership, depending on the organisation's maturity and objectives.

 \item \textbf{Oversee OSS Compliance} - OSPOs manage compliance with OSS licenses to mitigate legal risks. This includes designing processes and policies to ensure obligations are met, protecting intellectual property, and managing potential infringements.

 \item \textbf{Establish and Improve OSS Policies and Processes} - OSPOs develop policies that guide the organisation's OSS activities, including consumption, contribution, and business decisions. They also create processes for project creation, code release, communication, and training.

 \item \textbf{Prioritise and Drive OSS Upstream Development} - Organisations must prioritise which OSS projects to invest in. OSPOs help identify critical projects and establish strategies for community engagement, ensuring alignment with the overarching OSS strategy.

 \item \textbf{Collaborate with OSS Organisations} - OSPOs engage with relevant OSS organisations and foundations to align with business goals and influence key technologies. This collaboration can provide strategic advantages and influence within the OSS community.

 \item \textbf{Track Performance Metrics} - OSPOs measure the impact of OSS activities through performance metrics that align with business goals. This requires expertise in both metric design and operationalisation to support strategic objectives.

 \item \textbf{Implement Inner Source Practices} - Inner Source involves using OSS development practices internally to improve collaboration. OSPOs facilitate this by providing infrastructure, education, and support for cross-functional teamwork.

 \item \textbf{Grow and Retain OSS Talent Inside the Organisation} - Training programs are essential for developing internal OSS capabilities. OSPOs offer workshops, mentorships, and ambassador programs to nurture talent and integrate OSS contributions into performance reviews.

 \item \textbf{Provide Advice and Support on OSS} - OSPOs advise stakeholders on leveraging OSS in business strategies, evaluating technologies, setting community engagement goals, and ensuring compliance in various contexts such as mergers or outsourced development.

 \item \textbf{Manage Open Source IT Infrastructure} - OSPOs manage the infrastructure necessary for engaging with OSS projects and supporting Inner Source initiatives. This includes tools for compliance tracking, vulnerability scanning, and contribution monitoring.
\end{itemize}

\begin{figure*}
 \centering
 \includegraphics[width=0.7\linewidth]{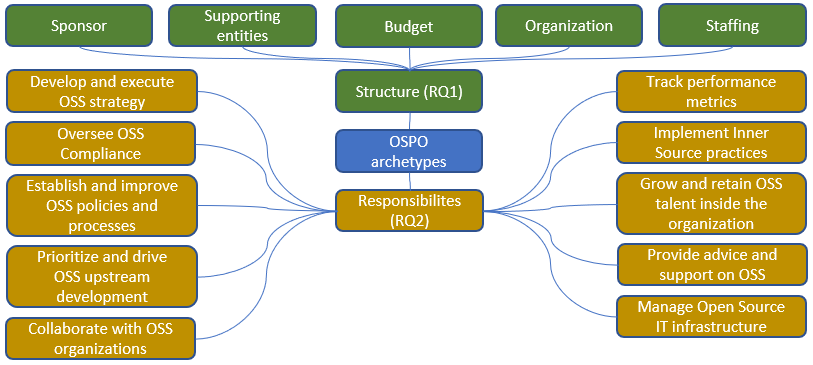}
 \caption{Overview of the conceptual framework based on the related work~\cite{haddad2020ospos}. Structure-related categories are explained in Section~\ref{subsec:OSPO-structures}, and responsibility-related categories in Section~\ref{subsec:OSPO-responsibilities}.}
 \label{fig:framework}
\end{figure*}

\subsection{OSPOs from a public policy perspective}
A report from OpenForum Europe~\cite{openforumeurope2022ospo} makes the case that the emergence of OSPOs is a result of the acknowledged need to increase OSS competence within different parts of the public sector administration. OpenForum Europe finds, looking at the trends from a managerial perspective that the motivations for building OSPOs in the public sector are diverse, but observe generally that there is a growing acknowledgement that OSS is essential for the public sector and that it is already used in everything from IT infrastructure to e-government services. 

Herpig~\cite{herpig2023ospos} specifically highlights the criticality of cyber security and the need for establishing a Cyber Security OSPO, a recommendation also highlighted by the US Cybersecurity and Infrastructure Security Agency’s (CISA’s) Open Source Software Security Roadmap~\cite{cisa2023ospo}. Main tasks include monitoring the national OSS supply chain, educating on OSS security practices, advising policymakers, supporting OSS infrastructure development, and interfacing with the broader OSS ecosystem. These recommendations are in line with the overall changes in the political environment of OSS and OSPOs where there is an increasing emphasis on software supply chain security, and the ability to make technical sourcing and design decisions based on local needs, values, and laws~\cite{ofe2022strategic}.


From a more general perspective, both OpenForum Europe~\cite{openforumeurope2022ospo}, and the Linux Foundation~\cite{todogroup2024survey} puts emphasis on how sharing and re-use of software between PSOs holds the biggest promise for potential benefits. Economic growth~\cite{blind2021impact}, increase in competitiveness and entrepreneurial growth~\cite{nagle2023government, wright2023open}, and long-term interoperability~\cite{ec2017interoperability} make up some of these benefits, as well as targets for potential policy interventions. 

Several reports, hence, also provide policy recommendations including the establishment of OSPOs looking beyond the specific topic of cyber security~\cite{herpig2023ospos, cisa2023ospo}. The Mission Bothorel~\cite{bothorell2021mission} report proposed the installation of an OSPO in the context of France, which was instantiated recently after in through the Free Software Unit at the French Interministerial Digital Directorate (Direction interministérielle du Numérique - La DINUM). Similar proposals has been laid forward in the US context~\cite{nagle2022strengthening}, proposing a federal government OSPO aimed at coordinating OSS policies other public sector OSPOs, as well as overseeing and advising in the general implementation of OSS policy interventions, and interfacing the general OSS ecosystem and other OSPOs on the international level.

\subsection{Policy implementations and support initiatives beyond OSPOs}
Looking beyond the construct of OSPOs, there has been iterative reporting on overlapping and aligning needs and initiatives. Below, we review common challenges and the organisational approaches highlighted as ways to address them.

\subsubsection{Challenges and barriers for OSS adoption}
Digital government research draws attention to both the opportunities~\cite{mergel2015open, mergel2016agile} and challenges~\cite{shaikh2016negotiating, van2015adopting} related to OSS and open collaboration in government and PSOs. Challenges needing attention can be generalised across several areas. Culture is highly cited, characterised by risk-averse and conservative procurement practices~\cite{persson2024soft}, with a preference for the status quo~\cite{rossi2006study} and existing solutions~\cite{deller2008open, koloniaris2018possibilities}. The need for customisation and integration efforts is also a common inhibitor~\cite{cassell2008governments, koloniaris2018possibilities}. IT departments have been characterised as both \textit{``sceptical and doubtful about the possibility of a radical change in the way they work and support their departments and the possibility that they would be requested to learn and support something totally new''}~\cite{koloniaris2018possibilities}. 

Long-term support and mandate from policy and executive management are also common barriers~\cite{silic2017open}, where long-term vendor relationships (and lobbying)~\cite{oram2011promoting} and unknowns, such as costs and technical risk, are reiterated in reports~\cite{cassell2008governments, silic2017open}. Internal communication and hierarchy can also contribute to slow (or non-existent) adoption within a PSO~\cite{cassell2008governments}. 

Limitations in knowledge and capabilities to consider OSS in acquisition and procurement are further iterated~\cite{cassell2008governments, deller2008open}. This follows a historical trend in the public sector of relying on outsourcing to maintain and scale the technical capabilities needed to meet current needs~\cite{marco2020outsourcing, cinar2019systematic}, a trend especially evident among local governments~\cite{persson2024soft}. This has led to a general lack of technical and software engineering expertise across the public sector, not just in OSS~\cite{borg2018digitalization}.

To fill this void, vendors and service providers are critical, while availability varies~\cite{deller2008open}, and are commonly seen as a barrier for OSS adoption~\cite{cassell2008governments, koloniaris2018possibilities}. PSOs are characterised as sceptical towards the support that OSS communities can provide~\cite{deller2008open, koloniaris2018possibilities}, why professional support is critical to promote the adoption of OSS in general~\cite{lungo2007experiences}

\subsubsection{OSPO-like means of addressing challenges and barriers}
The need to foster internal technical expertise and capabilities is called for at both national~\cite{oram2011promoting} and local levels~\cite{shaikh2016negotiating} of government. There is also an expressed need for internal champions and advocates, also referred to as boundary spanners~\cite{van2015adopting}, who can raise awareness of the opportunities and risks related to OSS and act as bridges to the broader OSS ecosystem and relevant stakeholders to enable trust and knowledge transfer. These champions can act as internal consultants and provide support throughout a PSO, as well as in those with active collaborations and partnerships~\cite{shaikh2016negotiating}. These expert and support functions align closely with the OSPO construct~\cite{ruff2022rise}, and with the expressed need from industry to grow organisational capabilities to support OSS operations aligned with business goals.

For local governments and PSOs with limited resources, the will and budget to create a dedicated OSPO may not be a priority. In these cases, pooling resources and creating collaboration structures make economic sense~\cite{marco2020outsourcing}, and are standard practices in general OSS collaborations~\cite{mahoney2005non}. Examples have been reported on in, for example, Danish, Swedish and Wallonian contexts, showing how local governments collaborate through member-driven formal~\cite{frey2023we} and informal~\cite{persson2024soft} associations, co-owned public service-providers~\cite{viseur2023communesplone}, or through resourceful PSOs acting as the primary sponsor and facilitator of the OSS projects~\cite{linaaker2025public}.

The literature stresses the need for professional support, as local government (even when organised) especially lacks technical capabilities. Cases (e.g.,~\cite{frey2023we}) stress the importance of \textit{``healthy ecosystem[s] of small and medium sized firms is stimulated to service public sector open source products''} that can ensure availability of \textit{``detailed documentation, quality assurance, certification schemes, training, and support services''}~\cite{shaikh2016negotiating}. The associations, as well as OSPOs, can act as an interface and facilitator of the dialogue and relationship between the concerned PSOs and vendors to monitor and support \textit{``the open source approaches, so that source code can be reused across the government''}, and that development follows in an open and agile practice~\cite{mergel2016agile}. Examples show that soft vendor lock-ins can easily become a reality~\cite{persson2024soft}.

\subsection{Conceptual framework}
Below the conceptual framework is presented (see Fig.~\ref{fig:framework}), which is used to characterise OSPOs inside PSOs and capture their various practices and challenges, but also to provide recommendations for PSOs in terms of how they may consider setting up their own OSPO. Specifically, it is expected that there will be a set of archetypes and responsibilities similar to the work by Haddad~\cite{haddad2020ospos}, yet with different characteristics from the industrial context.

\section{Research design}

\begin{figure}
 \centering
 \includegraphics[width=\linewidth]{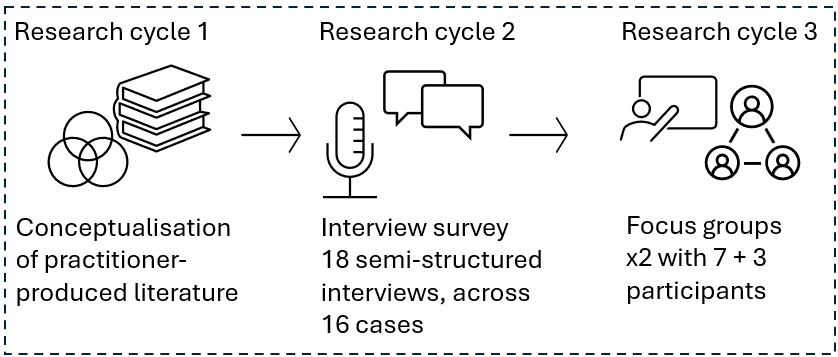}
 \caption{Overview of the research process over three research cycles, including i) conceptualisation of practitioner-produced literature, ii) a semi-structured interview survey, and iii) focus groups.}
 \label{fig:research-process}
\end{figure}

The study presents an exploratory qualitative survey~\cite{melegati2024qualitative} investigating how the adoption, development, and collaboration on OSS can be explicitly enabled through organisational support functions or centres of competency using the OSPO construct. More specifically, the study considers an OSPO as an institutional organisational construct that supports and accelerates the consumption, development, and collaboration of OSS (adapted from OpenForum Europe~\cite{openforumeurope2022ospo}.

It should, however, be noted that the OSPO label may be interpreted ambiguously and is not always labelled as such in PSOs. Hence, when searching for these support structures or functions, it was not expected that PSOs would have adopted the OSPO label. Accordingly, the unit of analysis is the support functions, independent of organisational structure or label.

With this scope, the following research questions were defined:
\begin{itemize}
 \item[RQ1:] How are support functions for OSS (i.e., OSPOs) organised and structured within PSOs?
 \item[RQ2:] What are the responsibilities of these support functions (i.e., OSPOs), and how are they performed?
\end{itemize}

Qualitative surveys are \textit{``research methods aiming to identify the diversity of characteristic values, rather than measuring their distribution, in a targeted population''}~\cite{melegati2024qualitative}. In this study, the investigation is conducted iteratively over three research cycles as illustrated in Fig.~\ref{fig:research-process}: i) synthesis of a conceptual framework based on industry practice to provide deductive foundation for empirical investigation; ii) semi-structured interview survey across cases representing public sector OSPOs to inductively address our defined research questions; iii) focus-groups to validate and further enrich interview findings. Below, we expand on these in more detail.

\subsection{Research cycle one: Conceptualisation of industry practice}
To create a conceptual framework for the empirical and inductive investigation of public sector OSPOs, we build on the work by Haddad~\cite{haddad2020ospos} which provides a comprehensive source representing the body of knowledge from a practitioner context. The source was selected through a purposive review of practitioner-produced literature (similar to grey literature) of OSPOs in the industry context where practice has been thoroughly reported. Specifically, we leveraged reports and material from the TODO group~\footnote{\url{https://todogroup.org/}}, and OSPO Alliance\footnote{\url{https://ospo-alliance.org/}} - two OSS communities gathering OSPOs mainly in the industry context. This focus follows from scientific reporting on OSPOs in general being sparse~\cite{harutyunyan2019industry, munir2021rise, xia2023lessons, riehle2024open}, and even sparser in the public sector context. 

Haddad's work has also adopted as the foundation for the TODO group's definition and classification of OSPO's structure, roles, and activities~\cite{todo2024mindmap}. This was validated through personal email communication with a board representative of the TODO group. An initial conceptual framework was created to highlight the structures and responsibilities of OSPOs in an industry context, providing a foundation for characterising OSPOs in the public sector. The framework is presented in Fig.~\ref{fig:framework}. The structure-related categories are explained in Section~\ref{subsec:OSPO-structures}, and responsibility-related categories in Section~\ref{subsec:OSPO-responsibilities}.

\subsection{Research cycle two: Semi-structured interview survey}
Informed by the conceptual framework from cycle 1, semi-structured interviews were conducted to empirically investigate the problem domain.

\subsubsection{Interview sampling}
The first step was to create an overview of the population to sample from. In this process, PSOs with established OSPOs were mapped within the 27 EU member states, plus Norway, Liechtenstein, and Iceland (an exercise conducted in spring 2023)\footnote{The EU-centric geographical limitation was given by the commissioning body. Norway, Liechtenstein, and Iceland were included because they are active members of the European Economic Area (EEA).} The following networks and resources were leveraged:

\begin{itemize}
 \item The EU OSPO network, facilitated by the European Commission OSPO
 \item OpenForum Europe's community network\footnote{\url{https://openforumeurope.org/open-source/}}
 \item The Open Source Observatory's Country Intelligence Reports\footnote{\url{https://interoperable-europe.ec.europa.eu/collection/open-source-observatory-osor/open-source-software-country-intelligence}}
\end{itemize}

The synthesised population (presented in table~\ref{tab:ospo-landscape}in Appendix C) consisted of 23 public sector OSPOs. This was not expected to be a complete representation of the population due to the aforementioned ambiguity in the construct, and the label may not always be used. Still, it provides a large enough population to sample from to gain a diverse set of cases (i.e., a public sector OSPO). 

From the identified population, 12 cases were initially selected using a purposive sampling approach~\cite{patton2014qualitative} to achieve diversity across i) level of government (local, regional, national), ii) nationality (west, east, mid, north, south Europe), iii) scope (goals and vision), and iv) maturity (size and age). The sampling was performed in collaboration with the study's commissioning body, which was interested in a diverse, broadly representative panel. To enable cross-analysis and provide context, cases were chosen within four initial clusters referring to local, regional and national levels of government, and Academia, with additional information collected as available and possible to further inform the sampling process. While our focus is on public sector OSPOs, we made an exception by including Code for Romania, a civil-society organisation that supports governments at all levels in Romania in adopting OSS solutions, thereby acting as an external OSPO-like function for Romania and providing an interesting case to compare and contrast with.

As these clusters evolved into what the study refers to as OSPO archetypes in section 4, four additional cases were included from the previously collected population (resulting in 16 included cases in total) to facilitate contrast and comparison within the different clusters. An exception was the Organisations with OSPO-like support functions archetype, which is only represented by the case of Code for Romania, as we did not find a comparable entity in our population. The decision not to include any further cases was influenced by both the availability and knowledge of potential corresponding cases, as well as the limited time and resources available for the execution of this study. A brief description of each case is provided in Appendix A.

\subsubsection{Data collection}
For each case, 1-2 interviews were conducted with key individuals, resulting in a total of 17 interviews with 18 individuals. Two representatives from the French Public Employment Service's OSPO were interviewed in the same session, and two from the European Commission's OSPO at separate occasions. As this is an interview survey aiming for breadth rather than depth, we limited ourselves to only one interview per case, except for the two aforementioned cases where the case organisations themselves encouraged the inclusion of two interviewees to cover appropriate perspectives. It may be noted, however, that all OSPOs were limited in size, and our sampled individuals maintained key positions, typically with overall responsibility of concerned OSPOs.

A comprehensive overview of the interviewees is provided in Table 3 below. The interviews were semi-structured, allowing for digression in topics as the interviews progressed. This flexibility allowed us to capture additional nuance, context, and examples that might have otherwise been overlooked. The interview questionnaire (see Appendix B) was based on the conceptual framework derived from practice in research cycle one and primarily consisted of open-ended, exploratory questions aimed at addressing the defined main research questions. Questions under section B.1. connects to RQ1, and questions under B.2-11 relates to RQ2, and more specifically the structural attributes and responsibilities defined in our initial conceptual framework as defined in Fig.~\ref{fig:framework}.

The interviews lasted approximately 1 hour and were conducted and recorded online via Teams, Google Meet, Zoom, or Jitsi, pending the preference of the interviewee. The interviews were conducted by two researchers: the lead researcher led the interviews, and the second researcher took notes and asked follow-up questions as needed. The recordings were automatically transcribed using either built-in or third-party services. These transcriptions were then cleaned and processed to ensure readability and accuracy. Processed transcriptions were sent to the respective interviewees, along with a summary of the main findings, to validate the accuracy and make any necessary revisions or retractions.

\begin{table*}[!t]
\caption{Overview of the interviewees and focus group participants, their role, the OSPO they represent, and the country where the OSPO is based (CR), and the year they were founded (F).}
\label{tab:interviewees}
\begin{tabular}{p{1cm} p{3.5cm} p{8.5cm} p{1cm} p{1cm}}
\toprule
\textbf{ID} & \textbf{Interviewee role} & \textbf{OSPO entity} & \textbf{CR} & \textbf{F} \\ \midrule
I1-DIN & Free Software Officer & Free Software Unit at The French Interministerial Digital Directorate (Direction interministérielle du Numérique - La DINUM) & FR & 2021 \\ \midrule
I2-DEVIT & Software Engineer & Developers Italia at the Department for Digital Transformation (Dipartimento per la Trasformazione Digitale) & IT & 2017 \\ \midrule
I3-ZEN & Policy Officer & Centre for Digital Sovereignty (Zentrum Digitale Souveränität) & DE & 2022 \\ \midrule
I4-LUX & Senior Advisor & Luxembourg House of Cybersecurity under the Ministry of the Economy (Ministère de l’Économie) & LU & 2022 \\ \midrule
I5-EC & IT Project Officer - OSS Strategy & OSPO at the Directorate-General for Digital Services (DIGIT), European Commission & - & 2020 \\ \midrule
I6-EC & Project Leader - Open Source Software Activities & OSPO at the Directorate-General for Digital Services (DIGIT), European Commission & - & 2020 \\ \midrule
I7-POLE & Innovation Manager & IT Department at the French Public Employment Service (Pôle emploi) & FR & - \\ \midrule
I8-POLE & IT Manager & IT Department at the French Public Employment Service (Pôle emploi) & & \\ \midrule
I9-TAX & OSS Specialist & Bureau Open Source Software (BOSS) of Dutch Tax and Customs Administration (Belastingdienst) & NL & 2021 \\ \midrule
I10-CODE & Vice President of Product & Code for Romania & RO & 2016 \\ \midrule
I11-OS2 & Chief Executive and Secretary & OS2 – Public digitalisation Network (OS2 – Offentligt digitaliseringsfællesskab) & DK & 2012 \\ \midrule
I12-VNG & OSS specialist & Dutch Association of Municipalities (VNG - Vereniging van Nederlandse Gemeenten) & NL & 2019 \\ \midrule
I13-OC & Chief Executive Officer & Open Cities (Otevřená města) & CZ & 2016 \\ \midrule
I14-PAR & Open Source Programme Officer & IT department at City of Paris & FR & 2018 \\ \midrule
I15-BRAT & Chief Innovation Officer & Department for Digital Services and Innovation, at City of Bratislava & SK & 2019 \\ \midrule
I16-VENT & Deputy Executive Director & Digital Centre at the City of Ventspils & LA & 2003 \\ \midrule
I17-TRIN & Research Commercialisation Manager & Technology Transfer Office at Trinity College Dublin & IR & 2020 \\ \midrule
I18-LERO & Director of Lero OSPO & OSPO at Lero - the Science Foundation Ireland Research Centre for Software & IR & 2020 \\ \midrule
I19-AUST & Product-Manager Artificial Intelligence & Austrian Federal Computing Centre & AU & - \\ \midrule
I20-MUN & Open Source Program Officer & OSPO at City of Munich & DE & 2023 \\ \midrule
I21-ECH & Director of Digital Strategy & City of Echirolles & FR & 2021 \\ 
\bottomrule
\end{tabular}
\end{table*}

Interviewees were provided with upfront information about the study's background, design, and publication process, as well as the data collection and management processes used. They were offered the opportunity to ask clarifying questions about the process throughout the study. Each interviewee remains anonymous by name but is referred to by a generalisable and appropriate title.

\subsubsection{Data analysis}

Statements in the transcripts were coded in a stepwise sequence during the data analysis phase. Initially, structural coding~\cite{saldana2021coding}, a first-cycle coding method, was applied to organise the text data from the transcripts based on an a-priori code book consisting of the codes relating to the conceptual framework elicited in cycle 1:

\textbf{Structure-related codes (RQ1):}
    \begin{itemize}
        \item Goals of the OSPO: Benefits, value, and defined business and policy goals or objectives for OSS.
        \item Organisation of the OSPO: Organisation, internal sponsor, staffing, and budget of the OSPO.
        \item Supporting entities: Main stakeholders, entities, or individuals that are supported internally or externally by the OSPO.
    \end{itemize}
\textbf{Responsibility-related codes (RQ2):}
    \begin{itemize}
         \item Develop and Execute OSS Strategy
         \item Oversee OSS Compliance
         \item Establish and Improve OSS Policies and Processes      
         \item Prioritise and Drive OSS Upstream Development
         \item Collaborate with OSS Organisations
         \item Track Performance Metrics 
         \item Implement Inner Source Practices
         \item Grow and Retain OSS Talent Inside the Organisation
         \item Provide Advice and Support on OSS
         \item Manage Open Source IT Infrastructure
    \end{itemize}

Structure-related codes are further defined in Section~\ref{subsec:OSPO-structures}, and responsibility-related codes in Section~\ref{subsec:OSPO-responsibilities}.

The structured data was then coded using open coding, another first-cycle coding method used to inductively capture emerging trends or concepts in qualitative data~\cite{saldana2021coding}. Transcript excerpts were accordingly assigned inductive codes along with short descriptive notes summarising the main findings in the statements. The open codes and notes were continuously grouped using an axial coding approach (a second-cycle coding method~\cite{saldana2021coding}), and the a-priori codebook. The process rendered in the revised structural attributes (RQ1) and responsibilities (RQ2) presented in Fig.~\ref{fig:revisedFramework}. Finally, selective coding, a second-cycle coding method~\cite{saldana2021coding}, was used to further synthesise data across the codebook into six common archetypes:

\begin{itemize}
    \item National-Government OSPOs
    \item Institution-centric OSPOs
    \item Local government OSPOs
    \item Association-based OSPOs
    \item Academic OSPOs
    \item Organisations with OSPO-like support functions
\end{itemize}
Accordingly, each OSPO archetype is defined through its structural attributes (RQ1) and responsibilities (RQ2) as visualised in Fig~\ref{fig:revisedFramework}, representing our posteriori code book. The archetypes are further described in Section 4.

\subsection{Research cycle three: Focus Groups}
Focus groups were used to further validate and explore the OSPO archetypes and the synthesised analysis from the semi-structured interview survey~\cite{kontio2008focus}. 

\subsubsection{Data Collection and Analysis}
Two focus groups were planned to provide sufficient coverage and detailed validation of the different archetypes. The first focus group focused on the cases within the three OSPO archetypes: National Government, Institution-centric, and Academic OSPOs. Out of ten invitations, seven participants attended, including former interviewees (I2-DEVIT, I4-LUX, I5-EC, I6-EC, I17-TRIN) and an external participant (I19-AUST). The second focus group focused on the three remaining OSPO archetypes: Local Government, Association-based, and Organisations with OSPO-like support functions. Of ten invitations, three participants attended, including one former interviewee (I11-OS2) and two external participants (I20-MUN and I21ECH).

In each focus group, participants were presented with a background on the study, followed by overviews of the different OSPO responsibilities characterised by the relevant OSPO archetypes, as well as the challenges and drivers identified in the studied cases. After each overview, the research team facilitated an open discussion among the participants. Each focus group lasted for 90 minutes, was recorded, and transcribed. Transcripts were analysed using a similar coding approach used in the second research cycle. No new high-level codes emerged through the analysis, but served as a validation of existing findings.

\subsection{Threats to Validity}
Since a qualitative research approach was adopted, a set of criteria for naturalistic inquiries proposed by Guba~\cite{guba1981criteria} was utilised for discussing and evaluating the threats to validity of the study. These criteria include credibility, transferability, dependability, and confirmability.

\subsubsection{Credibility}
Credibility pertains to the truth value of the presented findings. To enhance this aspect, member checking was conducted, allowing each interviewee to correct, add, or retract any statements in the transcripts, as well as reviewing the full report before being published. Additionally, the researchers behind this study engaged in continuous discussions and peer debriefings to maintain situational awareness of their observations and the perceived level of saturation among the different OSPO archetypes. 

The follow-up focus groups further contributed to strengthening this aspect of validity. It may be noted that the second focus group only collected three participants. Yet, all three participants were experienced, and two were external and new to the data, why we consider their input as valuable and still a partial validation of the findings, beyond other efforts such as the member checking process. 

\subsubsection{Transferability}
Transferability concerns whether and how the presented findings can be applied to other cases beyond those studied. Since this is a qualitative and exploratory study, based on a limited sample of public sector OSPOs, the aim is not to make generalisations about an entire population. Instead, the goal is to develop an understanding of the problem context and generate knowledge that can be valuable for practitioners operating within this context. Consequently, readers must consider the context from which the data was collected to enable any anecdotal generalisations. Quotes from the interviews were used consciously to provide rich and contextually detailed findings.

Purposive sampling is another method that was employed to ensure the transferability of the results. The initial mapping yielded a broad sample of the overall population of PSOs with OSPOs (or OSPO-like functions) in place (see Appendix B). From this sample, sixteen cases were selected based on criteria aimed at achieving as wide a representation as possible.

\subsubsection{Dependability}
Dependability refers to the reliability of the results in terms of their replicability and traceability throughout the chain of evidence. To enhance this aspect, an audit trail was maintained throughout the data analysis and collection process. This documentation covers all stages of the research process, from the conceptual framework through the mapping, interview survey, and focus groups.

\subsubsection{Confirmability}
Confirmability relates to the extent to which the presented findings were objectively derived. In this regard, investigator triangulation~\cite{guba1981criteria}, i.e., the involvement of multiple researchers, was leveraged throughout the research design and implementation process to ensure neutrality and reduce the risk of introducing researcher bias. This approach includes several steps aimed at maintaining research focus and a clear chain of evidence, from the initial assumptions to the design of the questionnaire and the iterative coding process.

\section{OSPO archetypes}
\begin{figure*}
 \centering
 \includegraphics[width=0.7\linewidth]{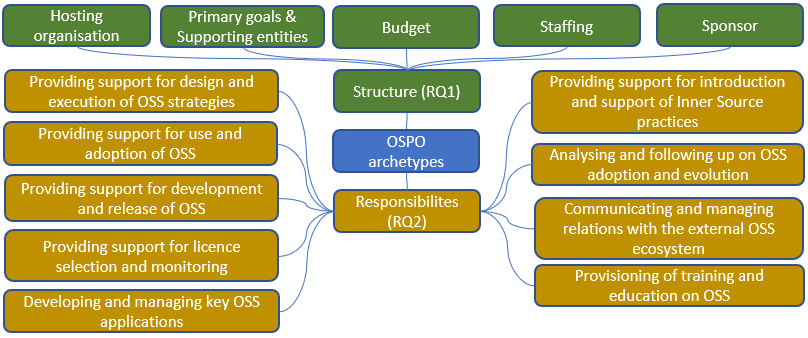}
 \caption{Overview of structural attributes (RQ1) and responsibilities (RQ2) of the OSPO archetypes based on the conceptual framework in Fig.~\ref{fig:framework}.}
 \label{fig:revisedFramework}
\end{figure*}

Our study investigates 16 cases of public sector OSPOs distributed across six different OSPO archetypes: National Government, Institution-centric, Local Government, Association-based, and Academic OSPOs. These archetypes, or categories, emerged organically through the investigation as additional cases were included. The archetypes can be further defined in terms of their structural attributes (answering RQ1), and responsibilities (answering RQ2), as visualised in Fig.~\ref{fig:revisedFramework}. The structural attributes are presented per archetype in Table~\ref{tab:archetype-overview}.

\begin{table*}[!t]
\caption{Overview of the identified OSPO archetypes based characterised based on conceptual framework from industry (see Fig.~\ref{fig:framework}.}
\label{tab:archetype-overview}
\begin{tabular}{p{2cm}p{14.5cm}}
\toprule
\textbf{OSPO archetype} & \textbf{Description} \\ \midrule

National government OSPOs & 
\textbf{Hosting organisation}: National administrations or ministries responsible for digital transformation and government (in general or for specific domains). 

\textbf{Primary goal and supporting entities}: Build and scale capacity in national public sector in adopting and collaborating on OSS.

\textbf{Budget:} Dedicated internal funding

\textbf{Staffing:} 3-70 FTEs (Management, Legal, Community, Engineering)

\textbf{Sponsor:} Executive management \\ \midrule
 
Institution-centric OSPOs & 
\textbf{Hosting organisation}: Internal departments responsible for IT service provisioning to the overarching institution.

\textbf{Primary goal and supporting entities}: Build and scale capacity inside institution in adopting and collaborating on OSS

\textbf{Budget:} Dedicated internal funding

\textbf{Staffing:} 2-4 FTEs (Management, Legal, Community)

\textbf{Sponsor:} Executive management \\ \midrule
 
Local government OSPOs & 
\textbf{Hosting organisation}: Departments responsible for IT service provisioning within local government (including municipalities, cities, and regions).

\textbf{Primary goal and supporting entities}: Enable adopting and collaborating on OSS in the digital transformation of the municipality

\textbf{Budget:} Dedicated internal funding

\textbf{Staffing:} 3-30 FTEs (Management, Legal, Community, Engineering)

\textbf{Sponsor:} Executive management/Local government politicians \\ \midrule

Association-based OSPOs & 
\textbf{Hosting organisation}: Associations with PSOs as members or owners

\textbf{Primary goal and supporting entities}: Enable members to initiate and collaborate on OSS projects addressing common needs

\textbf{Budget:} Membership fees

\textbf{Staffing:} 1-10 FTEs (Management, Legal, Community, Engineering)

\textbf{Sponsor:} Members (PSOs) through general council \\ \midrule

Academic OSPOs & 
\textbf{Hosting organisation}: Institutions for higher education and scientific research - Technology Transfer Office or hybrid

\textbf{Primary goal and supporting entities}: Provide support for dissemination of research outputs as OSS

\textbf{Budget:} Dedicated internal funding

\textbf{Staffing:} 2-10 FTEs (Management, Legal, Community)

\textbf{Sponsor:} Executive management \\ \midrule
 
Organisations with OSPO-like support functions & 
\textbf{Hosting organisation}: Organisations independent of ownership or membership of any PSO, for example within the civil society

\textbf{Primary goal and supporting entities}: Build and scale capacity in national public sector in adopting and collaborating on OSS

\textbf{Budget:} Dedicated internal funding

\textbf{Staffing:} 25 FTEs (Management, Legal, Community, Engineering)

\textbf{Sponsor:} Members (individuals) through general council \\ \bottomrule
\end{tabular}
\end{table*}

The latter two of these archetypes, the Academic and Organisations with OSPO-like support functions, to various degrees falls outside of the scope of this study which is to investigate OSPOs within the public sector. Regarding Academic OSPOs, these can be located both within private and public academic institutions. In this study, however, the sample is limited to public academic institutions. In terms of Organisations with OSPO-like support functions, this archetype falls outside of the public sector. Yet, the case and archetype is still included as one of its goals is to support the public sector, and it demonstrates an alternative model that is pursued in some jurisdictions, and the potential benefits and drawbacks of this approach.

The different OSPO archetypes are characterised in the subsequent subsections by comparing and contrasting the different cases of OSPOs in terms of the nine responsibilities as defined in Fig.~\ref{fig:revisedFramework}.

The responsibility of “Developing and managing key OSS applications” emerged from the production and co-development of critical applications needed by larger parts of the public sector beyond the focal PSO where an OSPO is hosted. This responsibility was especially apparent in the cases of Local Government and Association-based OSPOs. Also worth noting is that the focus, content, and presence of each responsibility will differ per OSPO archetype depending on the needs and goals of the OSPO, the level of government, its maturity, budget, and capabilities.

Another caution is that the presented list of OSPO archetypes is neither universal nor complete. The archetypes and their characterisation is based on the investigated cases and the output of the research approach presented in section 3. Variations and deviations of OSPOs will differ case by case, and certain types will have been missed. Also, the cases investigated vary in maturity and pace of evolution. Readers are recommended to consider the contextual factors per case and anecdotally compare and contrast with the target context.

\subsection{National government OSPOs}
National government OSPOs are hosted by PSOs, such as those responsible for matters related to digital government and transformation. These range in size and budget between smaller 2-4 people teams with a focus on general support, to larger 70+ people organisations which are more engineering-focused. Competencies typically include OSPO management, legal, community management, and engineering. In this study, four cases of such OSPOs were investigated: 
\begin{itemize}
  \item Free Software Unit at the French Interministerial Digital Directorate (Direction interministérielle du Numérique - La DINUM)
  \item Developers Italia at the Department for Digital Transformation (Dipartimento per la Trasformazione Digitale)
  \item Centre for Digital Sovereignty (Zentrum Digitale Souveränität)
  \item Luxembourg House of Cybersecurity OSPO at the Ministry of the Economy (Ministère de l’Économie)
\end{itemize}

\subsubsection{Providing support for design and execution of OSS strategies}
A central responsibility of National government OSPOs is to enable the operationalisation of high-level OSS strategies and policies determined by policymakers. The OSPOs act as mediators between legislative intent and practical implementation, translating legal or political mandates into actionable processes for PSOs. Through their coordinating role, OSPOs ensure that OSS is systematically considered in software procurement and development, while also providing guidance to support the drafting of further specialised OSS strategies within their remit, even if such tasks are less common due to varying degrees of adoption.

For example, in France, Italy, and Germany, national legislation and coalition agreements require PSOs to include OSS as a candidate when acquiring or procuring new software, and mandate that newly developed OSS is considered for release under OSS licences. The OSPOs, in turn, may support other PSOs within their scope to draft their own OSS strategies. This is yet an uncommon task, most probably due to the (in relative terms) limited adoption of OSS. 

\subsubsection{Providing support for use and adoption of OSS}
A common practice for National government OSPOs (noted across German, French, and Italian cases) is the establishment and maintenance of centralised knowledge-sharing platforms that serve as repositories and facilitators for OSS within the public sector. These platforms help to systematically index OSS solutions of relevance to and used by PSOs, ideally by leveraging standardised metadata standards such as public-code.yml\footnote{\url{https://yml.publiccode.tools/}} to ensure consistency and ease of use. As put by I3-ZEN:

\begin{quote}
    \textit{``We are often having the situation where things are developed in one end of Germany, and you're not aware of what's happening in the other end.''} - I3-ZEN
\end{quote}

Further, by codifying knowledge related to software selection, procurement, and adoption, National government OSPOs provide a critical resource that supports informed decision-making and lowers technical barriers. The Italian OSPO, for example, complements its platform with detailed procurement guidelines that assist public administrations in evaluating OSS projects from both technical and cost perspectives. In addition to the platforms and codified knowledge, OSPOs offer direct, capacity-dependent assistance to PSO engaging with OSS. This support is often limited to to resource constraints but is pivotal to use and interpret resources shared by the OSPOs.


\subsubsection{Providing support for development and release of OSS}
National government OSPOs play an important role as facilitators of OSS development and release within the public sector. They create integrated infrastructures that provide repositories for publishing OSS projects, and help to foster the conditions necessary for reuse and the growth of vibrant developer communities. The aforementioned knowledge-sharing platforms and software catalogues enable PSOs to contribute, discover, and collaborate on OSS initiatives. Communication mechanisms, ranging from integrated chat platforms to dedicated code hosting environments, serve as essential enablers for building and sustaining the social and technical collaborations fundamental to the success of OSS projects. Complementary to technological infrastructure, national OSPOs provide guidance through documentation, standards, and practical support, ensuring consistency and quality in OSS releases.

For example, the German OSPO’s knowledge-sharing platform includes a dedicated GitLab instance where government entities can host and share their OSS projects, actively enabling project visibility and collaboration. Across Germany, France, and Italy, OSPOs augment this infrastructure with explicit documentation and guidelines covering OSS project release, licensing choices, and artefact quality assurance. Beyond platform provisioning, the OSPOs offer tailored support to assist organisations in navigating the release process, ranging from recommendations on publication venues to hands-on review of licencing and technical compliance.



\subsubsection{Providing support for licence selection and monitoring}
A fundamental responsibility of National government OSPOs is to provide structured support for the selection of appropriate OSS licences, as well as for interpreting and managing the obligations these licences imply. OSPOs function as stewards of legal compliance, maintaining curated lists of recommended and discouraged licences, and offering regular updates based on evolving organisational requirements and incoming queries. This legal facilitation extends to the provision of procurement guidelines detailing the distinctions and implications of various licence types. Additionally, OSPOs implement technical measures, such as the requirement to generate Software Bill of Materials (SBOMs), which foster automated compliance and security evaluations of OSS projects prior to release. These SBOMs underpin efforts to enhance transparency, risk detection, and ongoing licence management, though such automation is often still emerging across national contexts.

For example, the German OSPO continually updates its list of recognised OSS licences in response to user needs, and requires SBOMs for all catalogue-listed software to enable automated and reliable compliance checks. The Italian OSPO, meanwhile, provides explicit legal guidance within its procurement materials, outlining how licence selection impacts both the use of external OSS components and the obligations when releasing new software projects. Across these offices, the capacity to deliver advanced compliance support varies, with automation and dedicated tooling representing ongoing areas of development.



\subsubsection{Developing and managing key OSS applications}
A special characteristic among some National government OSPOs is the assumption of stewardship and development responsibilities for strategic OSS applications that serve broader sovereignty or public infrastructure needs. In this role, the OSPO transitions from a supporting to  becoming a leading actor, taking charge of developing, integrating, and maintaining essential digital services and platforms. 

For instance, the German OSPO is tasked with enabling digital sovereignty for PSOs, and manages the ongoing development of OpenDesk, a digital sovereign workspace. This suite integrates mature OSS projects like Jitsi, Element, Nextcloud, and Collaboffice to deliver a complete set  of fundamental digital capabilities for the general civil servants. Contributions are made upstream to lead developers or vendor organisations, ensuring mutual benefit and long-term sustainability. Similarly, the House of Cybersecurity in Luxembourg has a long-established practice of developing OSS tooling for their own use cases, e.g., related to cyber threat management.

\subsubsection{Providing support for introduction of Inner Source practices}
Within National government OSPOs, formal support for Inner Source practices is generally limited or deprioritised. Instead, the OSPOs tend to favour public openness by default, encouraging development and collaboration that is accessible to the broader OSS community, rather than restricted to internal governmental circles. The rationale frequently cited is that initiating work as OSS from the outset minimises unnecessary duplication of processes and maximises the broader benefits and transparency associated with OSS development. As put by I2-DEVIT:

\begin{quote}
    \textit{``We have this opinion of saying, go open source, Inner Source is some unnecessary step. That will take you a lot of time with small benefits because potentially other contributors are likely to come from other administrations.''} - I2-DEVIT
\end{quote}

For example, while the French public sector does employ a GitLab instance to make internally developed software available across government entities, the national OSPO’s main focus is on enabling open collaboration with the public. The Italian OSPO, meanwhile, explicitly views Inner Source as superfluous, preferring that work be performed openly from the start to fully realise the advantages of the OSS model.


\subsubsection{Analysing and following up on OSS adoption and evolution}
An important function and oppotunity for National government OSPOs is to leverage the data generated via knowledge-sharing platforms and communication channels to support analysis and ongoing evaluation of OSS adoption and impact. However, the extent and sophistication of analytical activities vary considerably. While some OSPOs actively monitor metrics such as the number of registered PSOs and published OSS projects, others lack the resources to prioritise or institutionalise such analysis. Yet there is a broad recognition that robust metric tracking and follow-up can strengthen communication of impact and contribute to securing long-term operational funding (a challenge frequently cited).

For example, the French OSPO does not currently engage in systematic analysis or follow-up of OSS-related data due to resource constraints, despite seeing value in the practice. By contrast, National government OSPOs in Italy and Germany do track usage statistics within their catalogues, including metrics on organisational registration and project publication rates.


\subsubsection{Communicating and managing relations with the external OSS ecosystem}
Facilitation of technical communities is a generally adopted practice among the OSPOs that help to support the adoption and collaborative development of OSS in the public sector. The communities are composed of developers, designers, and wider stakeholder groups, and are aimed at fostering ongoing interaction, knowledge exchange, and project co-creation beyond the boundaries of individual PSOs. For example, the Italian OSPO coordinates the Developers Italia community, which includes over 10,000 participants from a variety of professional backgrounds, united around public sector OSS projects. In France, the corresponding community is called Blue Hats, which shares many similar features and characteristics.

This community-building effort can further be extended through advisory structures, which enable continuous dialogue between the public sector and experts from the surrounding OSS ecosystem. The French OSPO's free software council provides an example, consisting of representatives both from central administrations as well as from the general OSS ecosystem. The council meets regularly to discuss and provide input on topics of concern for the French OSPO or the council in general.

Cross-national communication and collaboration are also supported via dedicated EU OSPO network coordinated by the European Commission OSPO.



\subsubsection{Provisioning of training and education on OSS}
Provisioning of formal training and educational programmes is an important topic but often limited due to resource scarcity. The national OSPOs' primary approach typically centres on supplying practical guidelines and targeted support to PSOs. 

However, external communities provide complementing training activities in both cases, either facilitated or closely aligned with the OSPOs. In France, workshops are facilitated regularly through the Blue Hats, a community of civil servants and individuals interested in using and developing OSS in the public sector. The workshops help build and share knowledge across the community and its participants on general and technical application-specific topics. In Italy, online seminars are provided from time to time within the community, addressing different issues of concern for PSOs.


\subsection{Institution-centric OSPOs}
Institutional OSPOs are typically set up inside internal departments responsible for IT service provisioning within the institution, with the goal of building and scaling the capacity inside the institution in adopting and collaborating on OSS. These typically limited in size with 2-4 people teams  with competencies including OSPO management, legal, and community management. This study has explicitly looked at:

\begin{itemize}
  \item European Commission's OSPO at DG. DIGIT
  \item IT Department at the French Public Employment Service (Pôle emploi)
  \item Bureau Open Source Software at CTOs Office of Dutch Tax and Customs Administration (Belastingdienst)
\end{itemize}

\subsubsection{Providing support for design and execution of OSS strategies}
Within Institution-centric OSPOs, there is a strong focus on shaping, implementing, and communicating OSS strategies and policies tailored to the specific context and needs of their host organisations. These OSPOs actively engage in building awareness and trust around OSS adoption, balancing organisational opportunities with perceived risks to support the practical release and use of OSS. Communication of the strategy and continual input into its evolution are also central tasks, fostering a pragmatic and trust-based view of OSS internally. In more decentralised or informal arrangements, some OSPOs integrate OSS working methods as a core part of organisational culture or “DNA,” relying on established norms rather than formal OSPO structures.

For example, both the European Commission and the Dutch Tax and Customs Administration OSPOs are involved in providing input and shaping OSS strategies and policies, with a strong emphasis on internal awareness-building and trust. 

\begin{quote}
    \textit{``Everybody thinks open source is something we don't use, right? Well, we use it for 80, 90\% of our software development process. So you need to explain that you make it more explicit…. But having a policy in place would really help us to end the discussion of, well, there's no policy.''} - I9-TAX.
\end{quote}

This approach is contrasted by the French Public Employment Service, where open, commons-oriented work is embedded throughout the organisation, resulting in decentralised and informal governance rather than a formalised OSPO presence.

\begin{quote}
    \textit{``"It is kind of the DNA of [The French Public Employment Service] to work in the public and to contribute to common goods''} - I7-POLE
\end{quote}


\subsubsection{Providing support for use and adoption of OSS}
Institution-centric OSPOs strive to balance between developer autonomy in adopting OSS and the implementation of structured processes to manage risk and quality. Developers are generally encouraged and empowered to independently select OSS components based on need, reflecting a culture of trust and responsibility. However, this autonomy is underpinned by clear intake processes, which may include vulnerability scans, health assessments, and periodic expert reviews, aiming to ensure the security, stability, and long-term viability of software choices within the institution. This practice is exemplified across all three cases.

Supportive infrastructure, such as test environments for safe OSS evaluation (exemplified in the European Commision), can further allow staff to assess OSS alternatives within a typically risk-averse IT culture.


\subsubsection{Providing support for development and release of OSS}
Institution-centric OSPOs are often liberal in granting developers autonomy in releasing and contributing to OSS, supported by lightweight approval processes focused on risk identification and compliance. As observed across the cases, while the encouragement of openness and contribution is central, internal controls, typically centred on security, intellectual property, and community management considerations, provide necessary oversight without stifling work. 

OSPOs may also implement mechanisms, such as periodic audits, to identify legacy code that could be made open, thus furthering transparency and reuse, a practice observed at the European Commission's OSPO. The French Public Employment Service asks whether:

\begin{quote}
    \textit{``“[the project is] useful for anyone other than [the French Public Employment Service], and in what way? And what is the volume? What is the number of users that could benefit from it?''} - I7-POLE
\end{quote}



\subsubsection{Developing and managing key OSS applications}

Institution-centric OSPOs are commonly positioned within internal IT departments of large PSOs, which themselves possess substantial development capacities and maintain multiple OSS projects. The OSPOs function primarily as support entities; rather than conducting development work directly, their role is to facilitate and guide the internal development and management of strategic OSS applications. By providing coordination, oversight, and capacity-building, the OSPO helps ensure the successful adoption and governance of OSS within its organisation.

\subsubsection{Providing support for licence selection and monitoring}
A recurring characteristic among Institution-centric OSPOs is the use of SBOMs to map the licences of OSS components used and released by their organisations. The extent and granularity of SBOM use varies, but they commonly serve as tools for internal license tracking and compliance, particularly when distributing OSS externally. 

For example, in the European Commission, SBOMs are used at the project level to track which OSS components and licences are present, though this does not extend to the full dependency graph or the entire Commission. Prior to OSS release, internal IP checks scan the codebase for licences and support recommendations that consider both the direct code and its dependencies. Practices vary across the French Public Employment Service and the Dutch Tax and Customs Administration, with SBOMs generated to different degrees for internal and external OSS projects.


\subsubsection{Providing support for introduction of Inner Source practices}
Inner Source practices are enabled or supported to differing degrees by the Institution-centric OSPOs, yet widespread implementation remains challenging for a range of reasons. Typically, technical platforms exist that allow internal visibility and potential collaboration, such as development environments where all staff can view and contribute to each other's code. However, there is often limited formal process or incentive structure to actively facilitate Inner Source programmes. OSS development is widely considered the ideal approach for software projects, with Inner Source treated as a secondary option suited primarily for projects that cannot be publicly released as OSS. Still, Inner Source is viewed as a valuable gateway for preparing software projects for OSS release and developing internal skills, something highlighted across the cases.



\subsubsection{Analysing and following up on OSS adoption and evolution}
Tracking, reporting, and using metrics to monitor OSS activities and support internal advocacy is considered important but varies in practice. Some OSPOs as the European Commission's, may establish formal processes for quantitative tracking and reporting to management, using these data to demonstrate progress, motivate continued investment, and provide oversight of OSS projects hosted on internal platforms. By contrast, neither the French Public Employment Service nor the Dutch Tax and Customs Administration have established metrics or processes for monitoring the evolution of their OSS operations, although some intake process data and development activity from platforms like GitHub can be accessed.


\subsubsection{Communicating and managing relations with the external OSS ecosystem}
Institution-centric OSPOs function as interfaces and liaisons, bridging internal organisational stakeholders and external actors across the OSS ecosystem. The OSPOs systematically connect individuals and teams within their host organisations to appropriate external communities, networks, and vendors, facilitating collaborations and knowledge exchanges that advance strategic goals such as procurement, sourcing, and project development. Engagement extends from local and national partnerships to European and global initiatives, ensuring continual alignment with wider OSS innovations and best practices.

For example, EU OSPO network (facilitated by the European Commission's OSPO) plays an important role in connecting that public sector OSPOs across Europe, but also internally inside the organisation.

\begin{quote}
    \textit{``“[The European Commission OSPO provides a] low barrier point of contact for anyone involved with the open source on the outside of the institution. On the inside, we are hopefully seen as a useful resource to get advice from when they look for open source communities, solutions they hadn't thought of, and generally for answers to any question related to open source.''} - I5-EC
\end{quote}

The French Public Employment Service conducts similar external collaborations, including involvement in TOSIT, a multi-sector community dedicated to joint development of OSS security infrastructure. In the Netherlands, the Dutch Tax and Customs Administration engages through the Dutch OSPO Knowledge Network, promoting experience sharing and best practices among local PSOs.



\subsubsection{Provisioning of training and education on OSS}
Training and education help to build awareness, competency, and support for OSS adoption across organisational levels, and provides an important activity for the OSPOs. These activities are particularly focused on communicating the existence, rationale, and benefits of OSS policy, especially targeting middle management and diverse departments to foster cross-organisational engagement. Training initiatives may be formal, informal, or collaborative, with efforts spanning from programme development to experiential learning and advocacy. Support for OSS is often embedded in wider efforts to encourage interaction, peer learning, and alignment with institutional goals.

For instance, the European Commission OSPO is engaged in developing and implementing training programmes at various stages, complemented by active outreach and events such as internal hackathons to promote awareness and collaboration. In the French Public Employment Service, the IT department places strong emphasis on educating other departments about OSS, with a decentralised approach to learning that includes peer review and strong advocacy by tech leads to support collective development of OSS understanding.


\subsection{Local government OSPOs}
Local government OSPOs are typically hosted within local government departments responsible for IT service provisioning, with the goal of enabling adoption and collaboration on OSS in the digital transformation of the local government along with any of its institutions and subdivisions. These range in size and budget between smaller 2-4 people teams with a focus on general support, to larger 30+ people organisations which are more engineering-focused. Competencies typically include OSPO management, legal, community management, and engineering. The three cases selected for further analysis within this archetype are:

\begin{itemize}
  \item IT department at City of Paris
  \item Department for Digital Services and Innovation, at City of Bratislava
  \item Digital Centre at the City of Ventspils 
\end{itemize} 

\subsubsection{Providing support for design and execution of OSS strategies}

Local government OSPOs typically display a diversity of approaches to OSSe strategy and governance, closely shaped by local context, available resources, and the interplay with higher-level policy frameworks. In some cities as those of Bratislava and Ventspils, explicit OSS strategies are absent, yet the OSPOs aligns its activities with broader digital or innovation policies adopted by municipal leadership, these often reference OSS either implicitly or explicitly as part of citywide objectives. 

In other localities, the OSPO works under the mandate of national law requiring both the consideration and release of OSS, embedding action plans within the city’s overall digital transformation agenda. In the case of Paris, the City's engagement with OSS even pre-dates the national law as the City has been actively working with and releasing OSS since 2001, when the e-service platform Lutece was open sourced.



\subsubsection{Providing support for use and adoption of OSS}
There is a general emphasis among Local government OSPOs on the need for demonstrating a solid business case and being able to show potential for cost-efficiency when considering a potential solution irrespective of it being OSS or proprietary. The OSPO representative from Bratislava notes that OSS is not, by default, cheaper, but that there are softer benefits that can be difficult to convey. 

For Bratislava, this can be an issue as sourcing decisions are managed by other, more conservative and risk-aversive departments, than the Department for Digital Services and Innovation, which makes up the City' OSPO. The OSPO representative of Venstpils also highlights the softer values, for example, that of being able to customise a solution to their own needs with their resources and, by extension, not being dependent on any external service providers. 

\subsubsection{Providing support for development and release of OSS}

Local government OSPOs, often small and developer-centric, prioritise the continuous development and release of OSS, routinely publishing their outputs openly. Their operations are characterised by direct involvement in custom software development, while also integrating external procurement when needed for specific solutions. 

In both both Bratislava and Ventspils, OSS requirements are integrated into procurement processes, with documentation and specifications published as part of tender activities to enable transparency and promote open alternatives. The Paris OSPO is mainly focused on supporting and promoting the development of the Lutece platform. However, there is an increasing focus and will among the City's developers on contributing back to the external OSS ecosystem as good OSS citizens.


\subsubsection{Developing and managing key OSS applications}
Local government OSPOs typically represent resourceful cities and municipalities, enabling them to take on a leadership in initiating and stewarding major municipal digital tools and platforms. The scale and influence of such platforms can fluctuate over time, often reflecting investment in community management and collaboration. These OSPOs aim not just to support in-house use, but also to foster wider adoption and lower barriers for other municipalities, sometimes extending hosting and services through mutual agreements or shared infrastructure models. Community growth and external engagement increasingly guide the strategic direction of these offices.

For example, Lutece, an OSS e-service platform launched by the City of Paris in 2001, has grown to encompass around 500 plugins and serves as a central digital platform for Paris, though external uptake has remained modest. Still, should there be a need the project is open for contribution:

\begin{quote}
    \textit{``“Members of the community can if the need isn’t shared, they’ll still be able to commit the code upstream and the City will set a review committee that will approve and merge the contributions.''} - I14-PAR
\end{quote}

The Paris OSPO has prioritised efforts to increase community involvement and facilitate wider municipal adoption by streamlining installation and support. In Latvia, the City of Ventspils not only develops OSS platforms for local use but also hosts and maintains solutions for neighbouring municipalities as part of a collaborative, shared responsibility approach to regional digital infrastructure.



\subsubsection{Providing support for licence selection and monitoring}
A common practice among Local government OSPOs is to establish clear licensing preferences and implement regular monitoring processes to support compliance and risk management for OSS adoption and release. Licensing choices are informed both by regulatory context and the need to maximise dissemination and reusability; monitoring relies on curated inventories of in-use software, allowing timely responses to emerging risks and vulnerabilities. These practices often involve coordination between OSPOs, IT departments, and external partners such as auditors.

For example, the City of Ventspils recommends the EUPL licence for OSS due to its availability in Latvian, and conducts annual audits, alternating between internal and external auditors, to survey OSS use in production. In Paris, a preference for permissive licences exists to facilitate broader adoption and sharing of municipal OSS. The Paris IT department, supported by the OSPO, maintains a curated list tracking each OSS component, including its version, dependencies, and architectural relationships. This comprehensive inventory enables both impact analysis and rapid response should vulnerabilities arise in critical components.



\subsubsection{Providing support for introduction of Inner Source practices}
Inner source practice is not directly supported or promoted within Local government OSPOs, typically due to their limited size and scope. For municipal OSS projects, such as Lutece and the e-service platform of Venstpils, OSS serves as the main model for share and reuse.

\subsubsection{Analysing and following up on OSS adoption and evolution}
Metrics and follow-up on OSS operations are important but not a prime concern for Local government OSPOs due to their small size and scope. Metric programs can, for example, help to inform planning, development and community management of key infrastructure projects (such as Lutece). Metric programs can also be leveraged to investigate and follow-up on end-user satisfaction of OSS solutions. The Bratislava OSPO, for example, work actively in defining and following up on existing and new KPIs, including satisfaction scores from residents related to the internally developed e-service platform. These are considered very important as they provide a sign of improving the general trust towards government IT services.

\subsubsection{Communicating and managing relations with the external OSS ecosystem}
Local government OSPOs commonly facilitate and nurture active communication and collaboration networks with other municipalities and governmental bodies within their respective countries, often as part of a broader service provision or intergovernmental cooperation. The OSPOs in the Cities of Bratislava and Ventspils, for example, maintain ongoing communications with other Slovakian and Latvian municipalities respectively, as they provide or develop services for these PSOs.

These connections enable sharing of knowledge, resources, and services aimed at improving digital offerings across the public sector. Additionally, international relationships often complement domestic collaborations, facilitating learning and experience exchange with cities recognised for higher levels of OSS maturity and innovation. For example, Bratislava’s OSPO further engages internationally with mature European OSS cities such as Paris, Amsterdam, and Barcelona. Likewise, the City of Paris utilises external networks strategically to gather insights, share experiences, and enhance its own OSS initiatives.


\subsubsection{Provisioning of training and education on OSS}
Education and training efforts are typically needed across several types of stakeholders in local governments, including procuement officers, IT service managers, politicians and end-users. The Ventspils OSPO, for example, provides support and continuous education for the users of their e-services. However, the main training provided has been towards the suppliers in raising the knowledge and awareness of OSS in relation to procurement.

\subsection{Association-based OSPOs}
This category concerns associations of PSOs, which provide a common legal and administrative entity where members can initiate and collaborate on OSS projects in a neutral setting addressing common needs. They also support the use and adoption of OSS solutions created by their members. These OSPOs range in size and budget between 1-10 people teams with a focus on general support, and in some cases also engineering-focused. Competencies therefore typically include OSPO management, legal, community management, and engineering. This study has looked at three instances of Association-based OSPOs: 

\begin{itemize}
  \item OS2 – Public digitalisation Network (OS2 – Offentligt digitaliseringsfællesskab)
  \item Dutch Association of Municipalities (VNG - Vereniging van Nederlandse Gemeenten)
  \item Open Cities (Otevřená města)
\end{itemize} 

\subsubsection{Providing support for design and execution of OSS strategies}

Association-based OSPOs are collaborative entities formed by pools of PSOs, often municipalities, that aim to provide a neutral, standardised administrative and legal framework for initiating, developing, and governing OSS projects. Their key characteristic is serving as collective spaces where members can align governance models, development practices, and procurement approaches to ensure consistency, scalability, and trust. These OSPOs frequently function as community managers, facilitating cross-sector collaboration and bridging municipalities with suppliers or market parties, thus helping overcome individual resource limitations and amplify common impact.

For example, in Denmark, OS2 operates as a common association for municipalities, offering standardised governance and development structures through which members collaborate on OSS projects throughout their lifecycle. Similarly, VNG (the Dutch Association of Municipalities) acts as a "man-in-the-middle," enabling dialogue and joint development between municipal members and market suppliers, while Open Cities provides a legal entity for cities and regions to initiate or host OSS projects collaboratively. In all three associations, strategies and focus areas are steered collectively by the members, with the OSPO serving as a facilitator ensuring the sustainability, neutrality, and effectiveness of their shared work.


\subsubsection{Providing support for use and adoption of OSS}
Association-based OSPOs primarily focus on supporting, maintaining, and promoting the OSS projects initiated by their member organisations, serving as centralised hosts and facilitators for collaborative software development. The scale and scope of these OSS portfolios reflect both shared priorities and the available resources of the association and its community. The OSPOs often play a key role in maturing projects, enabling collaboration to expand, and providing tangible platforms for community learning and demonstration of impact. Some also offer hosting infrastructure as an added service for supported projects.

For example, OS2 currently manages around 25 OSS projects for Danish municipalities and expects to expand with several new projects in the near term. VNG’s flagship project, Signalen, moved from initial development under the City of Amsterdam to collective ownership within the association, serving as both a testbed and proof of concept for cross-municipality collaboration and impact. Open Cities, meanwhile, hosts and facilitates six OSS projects, most prominently Cityvizor, a tool for visualising public spending that is widely adopted by Prague and other cities. Unlike OS2 and VNG, Open Cities also provides direct hosting services for its projects, further supporting municipal stakeholders in OSS adoption and utilisation.



\subsubsection{Providing support for development and release of OSS}

Supporting development, release and stewarding of common OSS solutions is a key part of the Association-based OSPOs' responsibility. Below, we expand on methods and practices used in more detail.

\subsubsection{Providing governance structures for members' OSS projects}

Association-based OSPOs commonly implement similar governance structures for both their central organisations and the OSS projects they facilitate, though with some localised adaptations. They are established as member-led entities, typically composed of municipalities (and regions in the case of OS2 and Open Cities), with ultimate accountability to their membership. At the centre of these associations is a dedicated core team, such as a Secretariat, that carries responsibility for day-to-day management, maintaining governance frameworks, and supporting the launch of new projects.

For example, within OS2, each hosted OSS project is governed by its own sub-community of OS2 members, along with a dedicated technical steering group and individual financing arrangements. VNG features a dual structure, combining a general steering group to steer finances and broad strategy, with a separate technical steering group tasked with project planning and roadmap execution. In both OS2 and VNG, part-time product owners, based within active member organisations as funding allows, act as coordinators, collaborating closely with technical steering committees to oversee project activities and ensure cohesion across project stakeholders.



\subsubsection{Providing support for procurement and collaboration with external service suppliers}

Association-based OSPOs typically adhere to established public procurement frameworks, supporting their members by providing standardised templates, repeatable processes, and reusable catalogues of requirements to facilitate efficient and transparent sourcing for OSS projects. Procurement may be managed centrally within the association or by active member organisations involved in a specific project, allowing flexibility in approach. 

The associations maintain relationships with market parties, often formalised as non-binding memorandums of understanding with service suppliers, to foster collaboration aligned with governance and development values. OS2’s vendor ecosystem includes 65 suppliers, 10–15 of whom are consistently engaged for project work, underpinned by mutual understanding agreements regarding OSS governance. Development practices should strive to separate code creation from quality assurance to promote openness and mitigate risk (a practise used by OS2), though practical limitations may see these activities consolidated under a trusted vendor, with the community or designated product owners providing oversight as needed.




\subsubsection{Providing support for financing and scaling of members' OSS projects}

The Association-based OSPOs are typically enabled by a voluntary, member-driven funding model for OSS projects. The level of financial support for developing and maintaining each project depends on the number of PSOs that choose to join and share in the costs. This approach provides freedom and flexibility where any municipality can use or contribute to a project without obligation. Participation in joint funding, covering expenses proportionally, is however strongly encouraged and seldom resisted among active members. In OS2, 

\begin{quote}
    \textit{``“...anyone can start using [or] engage in a project - it’s open source. Joining the collaboration and paying a part of the expenses are voluntary. This is a core value and important freedom, plus it seldom happens that a municipality are not willing to join the community and pay their part''} - I11-OS2
\end{quote}

In both OS2 and VNG, project funding scales with community participation. In Signalen, around 15 municipalities currently contribute funds, covering a third of development costs (with Amsterdam sponsoring the remainder), but achieving longer-term sustainability is estimated to require backing from 30-40 municipalities, a pattern echoed in OS2 for certain projects. 

\begin{quote}
    \textit{``“OS2 also has projects that need volume to be sustainable. Meaning that there must be 20-40 municipalities for the expected costs to be covered at a fair price for each contributor. But there are also projects where 2-5 municipalities are enough''} - I11-OS2
\end{quote}

Open Cities operates on a similar principle, collecting membership fees that underwrite the development of shared OSS platforms, and offering access to services for non-members on a paid basis.

Three primary cost areas commonly arise for these associations: i) the core development and maintenance teams, ii) the engagement of a product owner, ideally situated within a member municipality, and iii) the management and compliance oversight of the code base (such as code reviews and regulatory checks).

\subsubsection{Providing support for licence selection and monitoring}
Association-based OSPOs typically adopt a clear licence policy for the projects they oversee, aiming to balance openness with ecosystem and vendor engagement. As in the case of OS2, a primary licence, such as MPL 2.0, may be selected to encourage commercial participation, though expectations do not always match practical outcomes. Flexibility is maintained to use other licences, such as GPL, where specific project requirements demand it. OSPOs also implement centralised compliance tools and standard practices to scan codebases for vulnerabilities, monitor inbound dependencies, and ensure legal alignment. In-depth compliance and code-base management are generally performed by core development teams within member organisations.


\subsubsection{Communicating and managing relations with the external OSS ecosystem}
Association-based OSPOs consistently engage with external OSS ecosystems, but each has its own strategic priorities and collaborations. Their external activity is shaped by a dual focus: amplifying local successes and fostering broader adoption and learning through national and international networks. Engagement serves multiple purposes: spreading their flagship projects, learning from external partners, raising the profile and credibility of their OSS initiatives, and connecting members across sectors for knowledge sharing and policy input.

For example, VNG is proactive in promoting cross-border collaboration and exporting its Signalen system abroad, while simultaneously investing in a national OSPO knowledge network that supports Dutch municipalities and PSOs, fostering peer learning and contributing to policy formation. OS2’s primary focus remains domestic, supporting Danish municipalities as its core membership, but it is active in international communities and conferences to stay current and share expertise, often with an eye toward collaborations like those with Sambruk in Sweden. Open Cities combines local project hosting with external engagement; its outreach efforts help build awareness and encourage adoption beyond its immediate membership.



\subsubsection{Provisioning of training and education on OSS}
A defining characteristic among Association-based OSPOs is the concerted emphasis on capacity building and knowledge transfer, particularly around legal considerations and the integration of OSS requirements into public sector procurement practices. Education is delivered both formally and informally, using workshops, conferences, and toolkits, and is typically tailored to the varying levels of readiness across member cities. The need for these initiatives is heightened by limited resources and competing priorities, which can impede progress and the uptake of more advanced approaches.

For example, Open Cities provides regular workshops and conferences, focusing especially on legal matters and practical guidance for procurement. Municipalities involved are often at different stages of maturity, and underfunding remains a persistent challenge. VNG is commonly invited by municipalities in transition phases, such as when replacing existing systems. Its role centres on bridging abstract governmental policy objectives like digital sovereignty with tangible collaborative efforts in OSS, an approach generally regarded as effective. Further, the National OSPO Knowledge Network, inclusive of VNG, the Kingdom and Interior Relations Department, and the Dutch Tax Office, underpins broader training and educational efforts. Here, work is underway to collate best practice toolkits for PSOs and suppliers, with the intent to share these resources on a public website to support knowledge dissemination and skill development.



\subsection{Academic OSPOs}
Academic OSPOs are generally hosted within an Institution focused on scientific research and/or higher education. These range are typically limited in size between 2-10 people teams with a focus on general support. Competencies typically include OSPO management, legal, and community management. This study has looked closer at: 

\begin{itemize}
  \item Technology Transfer Office at Trinity College Dublin
  \item OSPO at Lero - the Science Foundation Ireland Research Centre for Software 
\end{itemize} 

\subsubsection{Providing support for design and execution of OSS strategies}
A key characteristic of Academic OSPOs is their bridging function between university IPR policies, researchers’ commercial ambitions, and the principles of open science. Such OSPOs operate with a guiding philosophy that software outputs should be made as open as possible, but as closed as necessary, carefully balancing openness with obligations arising from contracts, funding arrangements, and inbound licensing. The OSPO actively supports researchers not only in interpreting these requirements, but also in navigating opportunities for spinning out companies and aligning business models with OSS practices.

For example, at Trinity College Dublin, the OSPO implements the College’s IPR policy by advising on whether and how research software can be released under OSS licences, given contractual and funding requirements. The OSPO’s stance remains broadly neutral towards OSS, yet promotes openness as a default wherever feasible. Similarly, the Lero OSPO originated with a remit to help researchers develop, collaborate, and share software artefacts in the open. Its present role has evolved, now encompassing the wider field of open science; the OSPO was accordingly renamed to reflect its function of enabling free and transparent access to scientific outputs as a whole. This broadened approach was prompted by external review and a growing awareness of the need to maximise the impact of public research through engagement with the OSS and open science ecosystems.



\subsubsection{Providing support for development and release of OSS}
Academic OSPOs have a dual function. One function, as in the case of Trinity College Dublin, the OSPOs provide clarity and direction on matters of intellectual property management and distribution, and supporting researchers in the grant writing process. Consideration also needs to be made to the requirements of the funding institutions. Some companies sponsoring a project might want to keep results proprietary, especially when the research output includes or is based on proprietary IPR. This may be an issue in cases where government funding is matched against company investments. In these cases, government funding cannot be seen as a state aid preferential competitive advantage for involved companies. OSS hence provides a good option for removing such risk of exclusivity of the developed IPR. Generally, though, OSS provides an easy and standard way of managing software-based IP and collaborating on its development. National and international companies typically prefer the OSS model before closed and non-standard contract arrangements.

A second function of Academic OSPOs, as illustrated by the Lero OSPO, is focused on supporting researchers in adopting OSS to develop, collaborate, and disseminate their software-based research artefacts. This is done mainly through workshops and mentorship provided by the internal community of OSS champions constituted by the OSPO. The Lero OSPO is, however, planning to help their TTO grow similar capabilities to that of Trinity College Dublin.

\subsubsection{Providing support for licence selection and monitoring}
Supporting the licence selection and providing input on the implication of different licences is a common practice for Academic OSPOs. The OSPOs educate and support the researchers on what licence alternatives there are and what implications they have specifically from a commercialisation perspective. The type of business model to be chosen for the spinout and requirements from funders play an essential role in the recommendation. If a spinout is initiated, a software bill of material will be requested to identify any inbound licences from upstream dependencies, which will further impact the choice of licence. These types of services falls natural in the case of Trinity College Dublin, where the OSPO is represented by the University's TTO.

\subsubsection{Analysing and following up on OSS adoption and evolution}
Metrics and evaluation can provide important insights on the amount and impact of research outputs disseminated as OSS, and help steer general open science efforts of the Universities. The Trinity College Dublin OSPO keeps track of relevant metrics but only on a tacit level, as cases are limited, and everything passes through the same individuals in the OSPO. The OSPO representative highlights that about 30\% of all collaboration agreements with industry partners use OSS licences for developed software, underscoring its importance in the spinout process. The Lero OSPO does not, at this point, systematically analyse the general OSS adoption within the institute. However, they recently performed an internal survey to identify the needs and pain points of the researchers.

\subsubsection{Communicating and managing relations with the external OSS ecosystem}
International networks such as OSPO++ are essential for connecting Academic OSPOs with other corresponding OSPOs, for example, in the US, where the universities have been at the forefront with the case of John Hopkins. Matchmaking enables valuable knowledge exchange and lessons learned between different OSPOs and ther overarching organisations in terms of OSS and open science practices. On the national level, both the Lero and Trinity College Dublin OSPO collaborate directly with each other, but also with local OSS evangelists and through the Open Ireland Network, the Irish national network for knowledge sharing on OSS and related topics.

\subsubsection{Provisioning of training and education on OSS}
A core function of Academic OSPOs is their role in advancing knowledge and practical understanding of OSS among researchers and students, as well as amongst university management. The OSPOs operate at multiple levels, raising awareness about OSS licensing, effective use of social coding platforms, and the legal and business implications of open sourcing research outputs. In the case of Lero, training is tailored, and delivered through a combination of one-to-one support, hands-on workshops, and targeted lectures, with a focus on equipping project teams and individuals to assess the appropriateness of OSS within their own work and navigate the wider institutional context.

At Trinity College Dublin, researchers are generally knowledgeable and open to releasing their software as OSS, but university management has been historically more reticent due to the involvement of public funds in spinout companies. The OSPO’s work thus heavily emphasises education on the commercial integration of OSS, providing both a practical resource and an institutional signal of support for open approaches. 

\begin{quote}
    \textit{``"The open source business models are the most important thing that we really talk about.''} - I17-TRIN
\end{quote}

The majority of training is provided directly to individual researchers and projects, supplemented by occasional classes for computer science students and new staff.

\subsection{Organisations with OSPO-like support functions}
Organisations with OSPO-like support functions have no formal tie to PSOs through ownership or membership. They support OSS's adoption, development, and collaboration through their own means. This study has investigated the case of Code for Romania, a not-for-profit civil society organisation without formal ties to PSOs and which is primarily funded through grants, donations, and sponsorships. 

\subsubsection{Providing support for design and execution of OSS strategies}
Being an independent and non-government OSPO, Code for Romania does not provide support on the design and execution of OSS strategies for any PSO. Their aim is rather to equip civil society and PSOs with OSS tools and solutions to enable them to provide better public services. Developed solutions should work and be at par with the requirements of institutional users. They see a general need for a core of systems needed by a more extensive set of PSOs, for example, regarding websites, registration systems, building registries, and complaint management. 

\subsubsection{Providing support for use and adoption of OSS}
Code for Romania generally encourages and promotes the use of internally developed OSS projects in collaboration with stakeholders and end-users. They today provide a set of about 50 different applications. These applications relate to areas such as education, citizen engagement, health care, the environment, and social services. All applications are maintained and offered as hosted services primarily for PSOs within Romania. If other countries are interested, they must run their own instances. 

\subsubsection{Providing support for development and release of OSS}
Code for Romania mainly considers external OSS projects as building blocks to their internally developed OSS projects. A primary reason is that they feel the externally developed solutions are generally not ready for production by a PSO, for example, in terms of quality, performance, and accessibility. Code for Romania works on putting the pieces together, raising the bar, and tailoring to public sector requirements. One exception mentioned regards Consul, a citizen-engagement platform originating from the City of Madrid is one example. Generally, any modifications or development of upstream projects are contributed back. 

\subsubsection{Developing and managing key OSS applications}
Code for Romania focuses on developing OSS tools and applications for needs addressing critical use cases in society, working closely with concerned stakeholder. The team has also developed a suite of about 30 OSS solutions targeting the humanitarian refugee situation that emerged due to the conflict in Ukraine. The solutions provide a common interface for the refugees entering Romania, for example, to apply for emergency housing, health services, women, and mother and child services specifically. Some solutions are managed by PSOs, for example, related to housing allocation. Others are managed by different civil society organisations designated with different responsibilities. Some of the solutions have been replicated internationally, including Moldova. In total, about 1.5 million refugees have been using the different services up until today.

Another notable example of an OSS-based solution regards a website builder for public institutions, which was initiated through research and collaboration with small and large municipalities to identify core needs. One key issue was that all municipalities needed a website. Often, the municipalities procured a solution and ended up in a lock-in situation, having to pay for smaller website updates. The website builder is now available in two versions, one for municipalities and one for national administrations, as they have different needs and regulations to abide by. The goal is to hand the solution over to the Agency for Digital Government, who in turn will support PSOs on different levels to adopt the tool. Code for Romania will, however, continue to develop and maintain the tool.

\subsubsection{Providing supporting for licence selection and monitoring}
All software developed is released primarily under the MPL2.0 licence. 

\subsubsection{Analysing and following up on OSS adoption and evolution}
Code for Romania works actively with metrics and has it as a core design principle on their checklist (inc. documentation and other OSS hygiene aspects). It enables them to validate their hypotheses and verify that the products work as intended. They also require that PSOs provide the necessary data when one of their solutions is being used. The data is kept on an aggregated and anonymised level. In one case, the collected data and overarching metrics helped identify certain instances of a housing allocation app among municipalities that were not working correctly.

\subsubsection{Communicating and managing relations with the external OSS ecosystem}
Code for Romania has extensive collaboration with other “Code4” organisations across the globe. Code for Romania is, however, considered the most advanced in terms of building solutions. All the Code4 organisations come together under an international organisation called Commit Global, where they share knowledge and collaborate across borders.

\subsubsection{Provisioning of training and education on OSS}
Code for Romania has a program focused on supporting OSS policies, connecting PSOs with experts, and training potential employees in government and existing staffers in parallel as they explore and develop new needs and OSS-based solutions for those.

The OSPO representative notes that the organisation is generally observing an increase in interest among policymakers and PSOs. Working together with stakeholders and users early on and showcasing solutions is considered to promote adoption and interest. Code for Romania is, however, not to be considered an advocacy group. The OSPO representative highlights that they do not ''stand in front of parliament'' but rather educate and talk to decision-makers and executives, proving the value and showing examples.

\begin{table*}[htbp!]
\caption{Synthesised activities and functions per OSPO archetype and area of responsibility}
\label{tab:responsibility-overview}
\begin{tabular}{p{1.3cm}p{2.2cm}p{2.2cm}p{2.2cm}p{2.2cm}p{2.2cm}p{2.2cm}}
\toprule
\textbf{Responsib.} & 
\textbf{National government OSPOs} & 
\textbf{Institution-centric OSPOs} & 
\textbf{Local government OSPOs} & 
\textbf{Association-based OSPOs} & 
\textbf{Academic OSPOs} & 
\textbf{OSPO-like Bodies} \\ \midrule

OSS Strategy & 
Translate national policy into practice; guide PSOs in drafting strategies & 
Shape and communicate institutional OSS policy & 
Align local actions to policy/law; integrate OSS into city strategy & 
Facilitate member-driven collective strategy & 
Advise on institutional policy and open science alignment & 
Identify and prioritise needs, provide solutions accordingly \\ \midrule

Use and Adoption & 
Maintain central platforms/catalogues; offer acquisition support & 
Support intake process; manage and monitor intake risk & 
Business case validation; enable customisation & 
Host software directories; liaise vendor support; enable adoption by members & 
Guide researcher choices; promote open science principles & 
Provide suitable solutions; support end-user adoption and operation\\ \midrule

Develop-ment and Release & 
Provide code collaboration platform; offer development/release support & 
Support release and contributions process & 
Develop and release in-house/procured OSS & 
Manage collaborative project portfolios; host infrastructure & 
Assist research software dissemination; support development and release & 
Develop new/customise existing OSS projects \\ \midrule

Licence Selection and Monitoring & 
Curated licence lists; legal guidance; procurement support & 
Compliance and SBOM automation; legal guidance & 
Inventory tracking; regular audits; procurement support & 
Preferred licence models; Compliance and SBOM automation; legal guidance & 
Advise based on project \& commercial requirements & 
Ensure compliance in OSS provided \\ \midrule

Stewarding Key OSS Applications & 
Steward and drive key OSS projects (e.g., OpenDesk) & 
Enable project maintainerships; Steward and drive key OSS projects & 
Steward and operate municipal OSS projects & 
Steward and maintain member OSS projects in collaboration with vendors & 
Steward key research tools; enable impact; promote cross-project funding & 
Steward and maintain key OSS applications initially, transfer with time \\ \midrule

Inner Source & 
Preference for OSS over Inner Source & 
Preference for OSS over Inner Source; enable internal collab. platforms & 
Primarily open by default; not applicable at scale & 
Facilitate internal collaboration among members & 
Promote collaboration through open science principles & 
Not applicable \\ \midrule

Metrics and Follow-up & 
Track platform/project use; monitor impact & 
Management reporting; track code hosting activity & 
Informal, situational tracking; limited formal analytics & 
Track member participation/project health & 
Tacit aggregation; impact monitoring & 
Monitor end-user and community activity \\ \midrule

External Relations & 
Build cross-sector communities; facilitate expert boards; EU OSPO network & 
Liaise between institution and OSS ecosystem; network engagement & 
Partner with other municipalities/cities; shared services & 
Convene cross-jurisdictional networks/event facilitation & 
Build research \& stakeholder networks on OSS research software & 
End-user (PSO) partnerships; Engage in existing OSS communities \\ \midrule

Training and Education & 
Provide guidelines/practical support; leverage community & 
Offer tailored support/training; peer review and advocacy & 
Knowledge transfer across functions; procurement guidance & 
Workshops/toolkits for members; collaborative training & 
Workshops; targeted mentoring programs and classes & 
End-user (PSO) education and training \\ \bottomrule
\end{tabular}
\end{table*}

\section{Discussion}
As in industry~\cite{haddad2020ospos}, we noted several ways in which OSPOs can be organised. The findings specifically revealed six distinct archetypes of OSPOs, each with its organisational characteristics and means of supporting OSS operations within the public sector. The archetypes vary and overlap both in structure and responsibilities (see tables~\ref{tab:archetype-overview} and~\ref{tab:responsibility-overview}, respectively). 

Considering \textit{OSS strategy and policy}, National government OSPOs act as mediators between policy and practice, translating national strategies into actionable support and guidance. Local and Institution-centric OSPOs adapt these strategies to fit local or organisational priorities, not always explicitly, but more abstractly in the context of digital transformation and innovation. The associations, in turn, enable collectives of PSOs to codify and share strategy, while Academic and OSPO-like entities are shaped by external review, research, and commercial objectives. In either case, the translation and mediation are critical to address the scepticism and conservative culture that commonly resides among political and executive leadership~\cite{silic2017open, cassell2008governments} and their long-standing vendor relations~\cite{oram2011promoting}.

In terms of \textit{use and adoption}, central catalogues highlighting OSS projects developed or in use by PSOs serve a central role in enabling identification and selection, along with procurement and acquisition guidelines, and other forms of documentation. These types of resources, together with communities of practice (e.g., BlueHats and Developers Italia), provide force multipliers in support that can be managed by OSPOs across the National, Local, and Association-based archetypes, where resources are scarce relative to the entities they are tasked with assisting. Here, procurement-related support is of specific importance~\cite{cassell2008governments, deller2008open} as the lack of knowledge and practice in navigating complex legal frameworks in relation to software and OSS acquisition is a recognised challenge~\cite{lundell2021enabling}, and can lead to lock-in~\cite{lundell2022effective} if not done properly~\cite{persson2024soft}. In the Institution-centric cases, which have more similar operations to industry contexts, the focus is more on enabling engineers to be autonomous in the intake of OSS components, while managing both legal and security-related risks that arise.

\textit{Development and release} of OSS also draws a line between user-oriented and engineering-oriented organisations, as well as the support required by each. Institution-centric OSPOs generally have well-defined processes, tools, and infrastructure to support their engineers, while National, Local, and Association-based OSPOs are more geared toward providing guidelines, high-level procurement-centric support, and common infrastructure such as national social coding platforms, a critical enabler for open collaboration~\cite{mergel2015open}. The latter group of OSPOs and their overarching organisations can still prove engineering-heavy, \textit{stewarding} and maintaining key applications for the public sector at large. Association-based OSPOs (similar to OSS foundations~\cite{mahoney2005non}) are especially well-suited as long-term bodies, as also exemplified in the literature~\cite{frey2023we, viseur2023communesplone, persson2024soft}. Still, examples across the national and local also show how single PSOs can take on a stewardship role~\cite{linaaker2025public}. Regardless of context, the Open rather than the \textit{Inner Source} way was a preferred means for collaboration.

Across both adoption and development, \textit{license compliance and legal support} plays a critical role, with various types of OSPOs providing similar support for tool automation and practical advice as needed. The former is more prominent in cases where engineering departments are present. The latter is more exclusively in scope, e.g., in Academic OSPOs, as they provide guidance related to commercial or project-related requirements and grant application writing.

\textit{Metrics and follow-up} are generally considered necessary, but practices are applied to a much lesser degree across the archetypes. Use cases include advocacy, signalling impact, and the need for continued funding. More resourceful organisations and OSPOs are more open to formal and tool-supported approaches, while those more resource-constrained rely on ad hoc reporting, tacit knowledge, or opportunistic internal surveys, with follow-up being less formalised.

\textit{Community management and outreach} is practised by most types of OSPOs, whether through national/international networks, industry, or other municipalities, recognising the importance of leveraging communities and open collaboration to benefit from many of the opportunities OSS can offer. Communities of practice, advisory councils, intra-PSO partnerships, and national and international OSPO networks are some examples. The OSPOs become both an interface and boundary spanner~\cite{van2015adopting}, enabling knowledge transfer and dissemination across an internal and external workforce.

\textit{Training and education} is also more or less practised across the various OSPOs as cases highlight a consistent and recognised need for maturing knowledge and capabilities, aligning with literature~\cite{marco2020outsourcing, borg2018digitalization}. The actual delivery, however, ranges from comprehensive, structured programs for well-resourced OSPOs to informal peer support or reliance on external communities in less resourceful cases. Topics of concern range from business models and open science principles (in case of Academic OSPOs), to procurement and governance (e.g., in case of Local government OSPOs). 

In general, \textit{deviances in support provided relate to the resource intensity, and observed need for OSS as a strategic means} in relation to the OSPOs' overarching organisations' general policy and business goals. National government, Institution-centric, and Association-based OSPOs exhibit the highest degree of structural professionalisation, with formal frameworks, standardised approaches, and sustainability planning, while local and Academic OSPOs show greater adaptiveness, granularity, and bespoke support, but also more direct capacity constraints. Organisations with OSPO-like support functions, such as Coding for Romania, coming from the civil society, demonstrate sectoral innovation and community translation but are usually not embedded in any formal PSO-driven governance and collaboration structures.

The \textit{type of support provided accordingly relates to the reliance of external resources}, which is much more significant for the PSOs within the local levels of government, in contrast to the more resourceful PSOs with the National government or Institution-centric OSPOs (aligning with others' observations~\cite{linaaker2025public}). Comparing the industry~\cite{haddad2020ospos}, the latter archetype is the one most in line with established templates, given its mainly internal focus and support from a typically large IT workforce in its OSS adoption. The National government OSPOs surveyed showed a higher emphasis on supporting other PSOs across various parts and levels of government.

Compared to the more organisation-centric focus of industry OSPOs~\cite{haddad2020ospos}, our investigation further shows, across the archetypes, that \textit{there is a will and need to support each other within the public sector in the collective adoption of OSS}, even though certain OSPOs may have more of an inward focus. This especially resonated on the local level of government, where, for example, municipalities have significant differences in capabilities and resources~\cite{frey2023we, viseur2023communesplone}. Those more capable fill an essential role in leading OSS initiatives and adoption~\cite{linaaker2025public}. They should be encouraged and supported to establish Local government OSPOs that, in turn, can act as accelerators and supporters for other local governments (including municipalities, cities, and regions).

In particular, \textit{Association-based OSPOs enable less capable PSOs, especially at the local government level, to collaborate and initiate OSS solutions by addressing everyday needs}. The associations act as OSS Stewards~\cite{persson2024soft}, providing a sustainable hosting and maintenance of OSS projects on neutral grounds similar to general OSS foundations and their often industry-driven OSS projects~\cite{mahoney2005non}. The associations further create a collective, more informed voice towards vendors, who are critical for both the adoption of OSS solutions~\cite{lungo2007experiences, koloniaris2018possibilities} and their sustainability~\cite{shaikh2016negotiating}. The associations (as other OSPOs) accordingly also provide quality assurance in verifying necessary OSS practices are used~\cite{mergel2016agile}, so lock-in can be avoided~\cite{persson2024soft}.

A common feature across the cases was also the \textit{leveraging of external communities and knowledge networks to exchange best practices, grow collaborations, and use as a force multiplier in the development of OSS projects} - again highlighting the role of OSPOs as boundary spanners~\cite{van2015adopting}. Civil society, volunteers and enthusiasts across sectors commonly played a key role in these communities. Civil society specifically fills a gap by enabling the use of OSS and helping to organise PSOs, both in the collaborative development of projects and towards a joint National government OSPO. PSOs should, hence, not only look to themselves but also to the rest of society for support and input. The OSPOs across the different archetypes can complement each other, forming a larger community where they collaborate and exchange resources and knowledge, building and scaling institutional capacity on a societal level.

\subsection{Implications for practice}
The study provides practitioners with knowledge and examples to use as a basis when considering and designing support structures to enable the implementation of OSS-related policies and strategies. Based on our findings, we provide the following recommendations for national governments:

\begin{itemize}
    \item National digital transformation agendas should formally incorporate OSS as a fundamental tool for realising policy objectives.
    
    \item The creation of National government OSPOs is essential to coordinate the implementation of OSS within national digital strategies and maximise their impact.
    
    \item PSOs should be assisted in aligning their own digital transformation strategies with national OSS ambitions, including the establishment of dedicated OSPOs to deliver on these goals.

    \item Cross-sectoral collaboration should be enhanced through the formation and facilitation of national-level public sector OSS and OSPO networks, aimed at fostering knowledge exchange and joint action.

    \item Comprehensive guidance should be developed, specifying circumstances and best practices for the consideration, acquisition, and creation of OSS by PSOs.

    \item Investments are needed to develop specialised education and training programmes to expand OSS expertise within the public sector workforce and among technology suppliers.

    \item Active participation from civil society organisations should be supported and encouraged, recognising their essential role in advancing public sector OSS adoption and collaboration.

    \item National OSS-related performance indicators should be integrated into broader digital governance frameworks to monitor progress and reinforce accountability in digital transformation efforts.
\end{itemize}

For Local governments (including Municipalities, Cities, and Regions), we make the following recommendations:

\begin{itemize}
    \item Local governments should articulate and adopt OSS strategies tailored to their specific objectives, local contexts, and available resources.

    \item The establishment of local OSPOs is recommended to ensure that OSS strategies are realised and embedded within municipal or regional digital agendas.

    \item Mechanisms should be established, such as associations or joint legal-administrative entities, for pooling expertise and resources across localities to jointly govern, procure, and sustain OSS-based projects.

    \item Larger or more resource-rich municipalities should adopt a leadership role in guiding and supporting smaller counterparts in leveraging OSS for digital change.

    \item Civil society involvement should be enabled and encouraged to enhance public sector capability and promote broader community engagement in OSS adoption and development.
\end{itemize}

For Higher Education and Research Institutions

\begin{itemize}
    \item Universities and research organisations should provide clear policy guidance on the strategic role of OSS in advancing their institutional missions.

    \item Researchers must be supported through infrastructure, advice, and training in using OSS for research collaboration, dissemination, and broader impact.

    \item Technology Transfer Offices are encouraged to promote OSS as a pathway for commercialising research and building sustainable business models around academic outputs.

    \item The role and value of OSS outputs should be recognised in research assessment, including in hiring, promotion, programme design, and funding allocation processes.

    \item Collaboration between Academic OSPOs and PSOs should be actively fostered, both nationally and across Europe.

    \item European research institutions should study and learn from international best practices in developing and scaling Academic OSPOs, to ensure global competitiveness and local impact.
\end{itemize}

\subsection{Implications for research}
The study has several limitations, including opportunities for future research. Among these is the generalisability and transferability of the presented results, which the reader must consider alongside the contextual factors at play in the various cases studied and how they together form a representation of the OSPO archetypes. 

The roles and responsibilities within the OSPOs will depend on the specific organisation and their own direction and goals for leveraging OSS, as defined by their OSS strategy or policy. By studying additional cases, future research can help clarify the nuances of the respective archetypes and how they may change, as well as their naming and the addition of other archetypes. Our broad sampling provides a foundation of knowledge for research to depart from in further investigating the role of OSS and open technologies at large as tools and vehicles for digital transformation and innovation in the public sector and governments.

Specifically, we encourage future research to explore:

\begin{itemize}
    \item the impact of different OSPO archetypes and practices on public sector digital transformation outcomes. Such research can shed light on which structures and processes deliver the greatest value in different contexts.
    \item how OSPOs translate national or organisational OSS policies and strategies into concrete action, and what kinds of policy instruments, goals, and incentive structures foster actual systemic uptake of OSS across the various parts of government and public sector.
    \item what factors enable some OSPOs to scale and professionalise their support, and what type of support proves most efficient across the OSPO archetypes and the context they support.
    \item other types of OSPO archetypes, for example, considering trans-national contexts with countries, lacking incentives and resources to develop their own support structures.
    \item how OSPO archetypes and practices differ (or not) in low- and middle-income countries, and what barriers and challenges, as well as opportunities and potential that OSS offers, compared to more developed contexts.
    \item how maturity and extent of support structures for OSS in government can be measured and followed up through digital government indices to help steer countries on a path towards achieving their policy goals.
    \item how OSPO archetypes and practices must adapt and consider other forms of open technologies, including open and shared data, open science, open hardware, and open source AI.
\end{itemize}

\section{Conclusions}
To sum up, we note that OSPOs come in many shapes and colors, and not always with the OSPO label. In general terms, they represent a support function, a policy enabler, and a change agent that drive of culture and organisational change, helping to address the fundamental fears and challenges implied by the openness, for example, in terms of viral licenses, usability, customisation needs, security concerns and lack support. The OSPO helps organisations acknowledge and manage these fears and challenges in a grounded and pragmatic way. The OSPO helps organisations to weigh the risks and costs against the benefits and potential of going open. The OSPO provide organisations with the necessary tools and introducing the best practices to leverage OSS as a tool to achieve their business or policy goals.

The latter is an important characteristic to highlight as establishing an OSPO without an overarching strategy or plan is of limited use. Equally, it is as meaningless to define an OSS strategy or policy without ensuring that there is sufficient knowledge, support, understanding, and infrastructure to implement the strategy or policy. In essence, the OSPO can be considered a policy enabler and support action, enabling an PSOs to to reach the different policy goals, be it sovereignty, interoperability or innovation. 

As a change agent, the OSPO would ideally dissipate with time as an organisation transforms and grows the capabilities required to leverage OSS accordingly, a phenomenon that has lately been observed in the IT industry . However, looking forward, we do not anticipate this trend anytime soon in the public sector. Rather, there is a need to widen the scope and focus beyond OSS, which can easily become siloed as that of open data. To achieve policy goals and the strive towards a more digitally sovereign, interoperable, and open society, scope and focus of policies and strategies, as of the OSPOs, need to include the data, the standards for how it is represented and communicated, the APIs between systems and their environments, the hardware it all runs on, and the science that drives the development forward. OSPOs will, hence, continue to play a pivotal role going forward, although the meaning of ''Open Source'' will most likely move beyond software towards open technologies in society at large.

\textbf{Acknowledgments: }
The European Commission’s DG DIGIT 
to OpenForum Europe and RISE through the OSOR project under Specific Contract 32 FWC DI/07929 with Wavestone SA, 
for which authors of this report are very grateful. The authors would further like to thank the interviewees and experts who shared their knowledge and experience underpinning the research findings presented in this study. The authors would also like to extend their gratitude to the many experts who have helped review the study through its different stages.

\appendix

\section{Case descriptions}
\subsection{National government OSPOs}
\subsubsection{France}
In France, the national OSPO is constituted by the Free Software Unit, comprising a team of three individuals, located at the French Interministerial Digital Directorate (Direction interministérielle du Numérique - La DINUM). The OSPO's primary focus is on supporting PSOs under the various ministries within the national level of government. They also have an implicit mandate to support the Directorate internally in terms of OSS. The OSPO was established in 2021 following a number of laws, decrees, and strategies that promote digital sovereignty, transparency, and trust in the public sector.

France's first Open Government Partnership National Action Plan (2015) commits the French government to promoting open government principles, including transparency, participation, and collaboration. In 2016, The French Law for a Digital Republic required PSOs to preserve the control, durability, and independence of their information systems. It also encouraged the use of free software and open formats in the development, purchase, or use of these systems (Article 9 of The Law ''pour une République numérique,'' sometimes called ''Loi Lemaire''). More directly linked to the formation of the OSPO, the so-called Bothorel Report~\cite{bothorell2021mission} recommended that the French government establish a national OSPO. The report argued that OSS is a key enabler of digital sovereignty and innovation, and that the government should do more to promote its use in the public sector.

\subsubsection{Italy}
In Italy, the OSPO is represented by the Developers Italia project, founded in 2017 and jointly managed by the Department for Digital Transformation (Dipartimento per la Trasformazione Digitale) and the Agency for Digital Italy (Agenzia per l'Italia digitale). The initiative stems from the requirements of the country's law on digital public administration dating back to 2012. Article 68 of the 'Code for the Digital Administration' instructs public administrations to consider software ''in the category of free and open-source software.'' The law now mandates public administrations to consider using OSS when procuring software solutions.

The OSPO's support extends across all levels of government. Within the Department, Developers Italia is managed by a smaller team within the Department's engineering group, focusing on providing technical support and services related to enabling OSS use and development. On the Agency for Digital Italy’s side, the focus is on providing legal advice and related support.

\subsubsection{Germany}
In Germany, the OSPO is constituted by the Centre for Digital Sovereignty (Zentrum Digitale Souveränität), a PSO currently under the Federal Ministry of the Interior since the end of 2022. The ambition, however, is to include different German states as owners in the future. Currently, the Centre for Digital Sovereignty is in a build-up phase, being piloted by a smaller team of individuals, with the goal of expanding to a size of 67 people in the near future.

The formation of the German OSPO can be traced back to a 2019 study by the German federal government highlighting the risks of depending on a few IT suppliers, emphasising increased costs and reduced autonomy. This concern was echoed in a joint paper by the federal Chief Information Officer (CIO), IT-Planning Council, and IT-Council, all aiming to enhance digital sovereignty. By 2022, the new coalition government's 178-page agreement, ''Daring More Progress,'' was released after extensive negotiations~\cite{sozialdemokratische2021open}. It emphasised the role of OSS in enhancing digital infrastructure and government services, considering it vital for digital sovereignty.

Furthermore, Germany's third national action plan (2021-2023) under the Open Government Partnership focused on OSS, mentioning initiatives like the Corona-Warn-App. The plan also detailed actions intended to promote the objectives of the cross-government initiative, including the establishment of the National government OSPO, the Centre for Digital Sovereignty.

\subsubsection{Luxembourg}
In Luxembourg, the National government OSPO may be described as consisting of two parts. In part by the House of Cybersecurity, under the Ministry of the Economy (Ministère de l’Économie), which is in the process of establishing its OSPO, and in part by a core team of experts within the Ministry. The intention is that the OSPO at The House of Cybersecurity will serve as a blueprint for PSOs in Luxembourg to follow and to improve on. The core team at the Ministry is providing input and documenting the blueprint and work progress.

There are no specific no structured proposals or policies for providing direction or guidance for OSS on the national level, beyond some encouragement provided through the General Block Exemption Regulation~\cite{thill2023open}. Digital sovereignty and cyber security are, however, considered as important policy objectives and targets why the House of Cybersecurity is a front runner in the country.

\subsection{Institution-centric OSPOs}

\subsubsection{European Commission}
The European Commission OSPO was initiated in 2020, and resides within the Department of Informatics (DG DIGIT), the Commission's internal IT service provider. The OSPO was formed on the basis of a recommendation of the European Commission’s OSS strategy defined in 2020~\cite{ec2020ossStrategy}. This strategy transitioned from being a Directorate-specific approach to a Commission-wide strategy, as it was formalised as a ''Commission Communication'' and endorsed by the College of Commissioners. The OSPO also champions OSS development across Europe. The OSPO today consists of a small team of about three Full-Time Equivalents (FTEs). The focus is on executing the European Commission's overarching OSS strategy and constituting a liaison between the European Commission’s departments, and external actors on OSS matters. Specifically, they support DG DIGIT and the other departments within the European Commission.

\subsubsection{The French Public Employment Service}
The French Public Employment Service (Pôle emploi) is France's national employment service and has an IT department of about 3000 people and 300 development teams. They have a decentralised OSPO in the sense that development teams have a generally high awareness and knowledge about OSS. Tech leads in each team provide expertise and advocacy for how teams should develop and collaborate on OSS. Internal policies and management explicitly encourage the use and development of OSS. An internal group of stakeholders, including users, lawyers, tech leads, and security experts, is to be formed which is aimed to gather once every month to respond to questions regarding what to contribute or release as OSS and how to work with and grow OSS communities. This group may, in some sense, be comparable to a formal OSPO.

\subsubsection{Dutch Tax and Customs Administration}
The Dutch Tax and Customs Administration (Belastingdienst) has an IT department of about 3500 developers and is a heavy user of OSS. Their OSPO is formally referred to as the Bureau of OSS and is hosted in the CTO's office of the IT department. While the Administration has a history of engaging strategically with OSS dating back to 2007 when the first internal policy was defined, the OSPO was only initiated in 2021. The OSPO currently consists of one full-time member but has defined space for four additional members. The OSPO is, however, supported by a virtual team of policy officers, lawyers, security officers, higher management, and enterprise architects on a needs basis.

\subsection{Institution-centric OSPOs}

\subsubsection{City of Bratislava}
The City of Bratislava is the capital and biggest City of Slovakia, with about 500 000 inhabitants. The City has 17 city districts under its umbrella, all functioning with some autonomy. The OSPO is driven by the Department for Digital Services and Innovation, headed by the City's Chief Innovation Officer since 2019. The department explores how digital innovation can improve the services provided by the City to its residents and civil servants. Currently, the department has 16 full-time employees with a combination of product and project managers, front- and backend developers, and DevOps professionals. In addition, there are about ten part-time employees with similar profiles.

\subsubsection{City of Ventspils}
The City of Ventspils is the sixth largest municipality in Latvia, with about 43 000 inhabitants. The OSPO is constituted by the Ventspils Digital Centre, founded in 2003, which is responsible for enabling and providing the City's digital services and infrastructure, including the operation of the City's own data centre, citywide optical data network, and implementing OSS-based solutions. The centre consists of 1st, 2nd, and 3rd-level support for technical operations. The main work is carried out by six system administrators within the Centre.

\subsubsection{City of Paris}
The City of Paris is the capital and largest City of France, with a population of about 2.1 million. It has 20 districts, or arrondissements, functioning with some autonomy. The OSPO, created in 2018, is constituted by one individual within the City's IT department, which has a staff of about 550 people. The department constitutes a support function for the rest of the City's departments.

\subsection{Association-based OSPOs}

\subsubsection{VNG}
VNG (Vereniging van Nederlandse Gemeenten) coordinates its member base of 342 municipalities on different aspects, including digital transformation. OSS and open standards are considered core pillars of that digital transformation. The VNG OSPO’ s main purpose is to enable the municipalities to collaborate on common OSS projects, for example, by providing the governance structures needed. The focus is currently limited to growing and maintaining the Signalen OSS project and its community. The OSPO, created in 2019, currently consists of one individual but with access to support from, for example, the marketing and communications team and the procurement team within VNG.

\subsubsection{OS2}
OS2 (OS2 – Offentligt digitaliseringsfællesskab) is an association created in 2012 today consisting of 80 out of 98 municipalities, but also including a smaller number of Regions and national public agencies. Membership is voluntary and commonly driven by a member PSO's IT, digitisation, or finance department. The main goal and purpose of the association are to develop and maintain a governance framework that members can work within to own and share IT solutions based on business needs. The association is facilitated by a central secretariat of four members plus nine part-time product owners, i.e., employees at the member PSOs who work part-time overseeing the technical planning and maintenance of specific OS2 projects. Member fees are used to pay wages and expenses of the secretariat. Each OSS project has its separate funding from the users of the project which is dedicated to development of the OSS.

\subsubsection{Open Cities}
Open Cities (Otevřená města) is a non-profit created in 2016 focused on helping cities, municipalities, and regions in their digitalisation, including the provisioning of OSS-based solutions. The general purpose is to collaborate on software needed by all members and to share the costs. They currently have 19 municipalities as members and two regions. A small number compared to the 6000 municipalities in the Czech Republic, however, due to the membership of large cities and regions, Open Cities concerns more than a third of the population in the Czech Republic. Today, Open Cities has about 12 employees, mostly part-time, of which six are developers, one DevOps-focused, project manager, business and members caretaker, marketing and events support and also six expert volunteers who are focused on Cyber Security.

\subsection{A5 Academic OSPOs}

\subsubsection{Trinity College Dublin}
The Trinity College Dublin OSPO, created in 2020, is constituted by the Technology Transfer Office, which generally supports researchers in the commercialisation process of their research, for example, securing funding, managing IP, and spinning out a company. The Technology Transfer Office spins out an average of 5-6 companies per year. The OSPO team comprises a business developer and a lawyer, as IP law is a pivotal part of the commercialisation process.

\subsubsection{Lero}
Lero, the Science Foundation Ireland Research Centre for Software, also created its OSPO in 2020. It is constituted by an informal committee with individuals across the institute in different ways knowledgeable of OSS. These individuals act as mentors supporting internal OSS-related efforts. A part-time director heads the community. The OSPO is currently in an early phase of being set up. A grant application has recently been submitted for external funding to help grow the OSPO further.

\subsection{Academic OSPOs}

\subsubsection{Coding for Romania}
This study has looked specifically at the case of Code for Romania, a civic tech non-profit launched in 2016, which today has about 25 individuals on their staff. The organisation is mainly financed through grants, donations, and sponsorships. No funding is received from any PSO. The organisation can compete on tenders but avoids them as they want to stay neutral and independent.

\section{Questionnaire for interviews}
Questions under section B.1. connects to RQ1, and questions under B.2-11 relates to RQ2, and more specifically the structural attributes and responsibilities defined in our initial conceptual framework as defined in Fig.~\ref{fig:framework}.

\subsection{Organisation and structure}
\begin{itemize}
    \item \textbf{Origin:} What motivated the creation of the OSPO?
    \item \textbf{Role:} Describe your role and responsibilities (in terms of supporting and enabling use, development and collaboration on OSS)?
    \item \textbf{Supported entities:} Who are the main stakeholders, entities, or individuals that you support internally or externally of your organisation? How are they organisationally connected?
    \item \textbf{Staffing:} What other roles are there that help in terms of supporting and enabling use, development and collaboration on OSS within your organisation or supported entities? How are you organisationally connected?
    \item \textbf{Sponsor \& budget:} Where in your organisation's hierarchy are you located? To whom do you (and any additional roles) report and get your budget allocated from?
    \item \textbf{OSS usage: } How does your organisation or supported entities use and leverage OSS today? What are the most prominent examples?
\end{itemize}

\subsection{Strategy}
\begin{itemize}
    \item \textbf{OSS Strategy}: How does OSS provide value for your organisation or supported entities? What are the overarching policy or business goals or objectives?
    \item \textbf{Policy \& Process}: How are strategic considerations defined, executed, and followed-up on within your organisation or supported entities?
    \item \textbf{Advice \& support}: What type of advice or support do you or your colleagues provide in this regard?
\end{itemize}

\subsection{Consumption}
\begin{itemize}
    \item \textbf{Policy \& Process}: How does your organisation or supported entities decide and manage what OSS is used? Are there any policies, or processes in place? If yes, how are these defined and operationalised?
    \item \textbf{Advice \& support}: What type of advice or support do you or your colleagues provide in this regard?
\end{itemize}

\subsection{Compliance}
\begin{itemize}
    \item \textbf{Compliance oversight}: How does your organisation or supported entities consider and address obligations implied by OSS licences?
    \item \textbf{Policy \& Process}: Are there any policies, or processes in place? If yes, how are these defined and operationalised?
    \item \textbf{Advice \& support}: What type of advice or support do you or your colleagues provide in this regard?
\end{itemize}

\subsection{Contribution}
\begin{itemize}
    \item \textbf{Policy \& Process}: How does your organisation or supported entities perform, manage and decide on contributions to external existing OSS projects?
    \item \textbf{Upstream prio}: How are engagement and contributions to OSS projects prioritised and supported?
    \item \textbf{Advice \& support}: What type of advice or support do you or your colleagues provide in this regard?
\end{itemize}

\subsection{Creation}
\begin{itemize}
    \item \textbf{Policy \& Process}: How does your organisation or supported entities perform, manage, and decide on the release of new OSS projects? Are there any policies, or processes in place? If yes, how are these defined and operationalised?
    \item \textbf{Advice \& support}: What type of advice or support do you or your colleagues provide in this regard?
\end{itemize}

\subsection{External relations}
\begin{itemize}
    \item \textbf{OSS org relations}: Does your organisation or supported entities have any relation to external organisations related to specific OSS projects, or OSS in general? If yes, what is the rationale, and how are these relationships managed?
    \item \textbf{Advice \& support}: What type of advice or support do you or your colleagues provide in this regard?
\end{itemize}

\subsection{Analytics, progress and follow-up}
\begin{itemize}
    \item \textbf{Metrics}: How does your organisation or supported entities track and follow-up on aspects such as risk, progress and impact related to how you use, develop, and collaborate on OSS?
    \item \textbf{Advice \& support}: What type of advice or support do you or your colleagues provide in this regard?
\end{itemize}

\subsection{Internal development}
\begin{itemize}
    \item \textbf{Inner Source}: Does your organisation or supported entities adopt (or strive to adopt) OSS development practices internally? If yes, please expand how.
    \item \textbf{Advice \& support}: What type of advice or support do you or your colleagues provide in this regard?
\end{itemize}

\subsection{Talent development, attraction, and retention}
\begin{itemize}
    \item \textbf{Training}: How is training and education considered or cared for in your organisation or supported entities related to how you use, develop, and collaborate on OSS?
    \item \textbf{Recognition}: In what ways, if any, does your organisation or supported entities reward or otherwise recognize or incentivise use, development, or collaboration on OSS?
    \item \textbf{Advice \& support}: What type of advice or support do you or your colleagues provide in this regard?
\end{itemize}

\subsection{IT infrastructure}
\begin{itemize}
    \item \textbf{Infrastructure}: What infrastructure and tool-support does your organisation and supported entities use to enable use, development, or collaboration on OSS, and any underpinning process or policy?
\end{itemize}

\section{Population of OSPOs in Europe}
The list presented in table~\ref{tab:ospo-landscape} presents an overview of public sector OSPOs in Europe (created in spring of 2023) from which interviewees were sampled as further explained in Section 3.
\begin{table*}[!t]
\caption{The OSPO Landscape in the Public Sector. }
\label{tab:ospo-landscape}
\begin{tabular}{p{2cm}p{2cm}p{4.5cm}p{7cm}}
\toprule
\textbf{Country} & \textbf{Type} & \textbf{Name} & \textbf{Host or Related PSO} \\ \midrule
Belgium & Regional gov & iMio & Government of Wallonia, over 300 municipalities \\ \midrule
Czechia & NGO & Otevřená města & Civil society org with government mandate \\ \midrule
Denmark & Regional gov & OS2 & Collaboration of Municipalities \\ \midrule
Denmark & Govt & Division for Technology and Data & Danish Agency for Digital Government \\ \midrule
EU & Govt & EU Commission's OSPO & European Commission \\ \midrule
France & Regional gov & Paris OSPO & City of Paris \\ \midrule
France & Govt & Free Software Unit & DINUM \\ \midrule
France & Govt & IT Department (Pôle emploi) & French Public Employment Service \\ \midrule
Germany & Govt & Centre for Digital Sovereignty (ZenDiS) & Ministry of the Interior \\ \midrule
Ireland & Academia & OSPO of Technology Transfer Office & Trinity College \\ \midrule
Ireland & Academia & LERO OSPO & University of Limerick \\ \midrule
Italy & Govt & Team Digitale & Government of Italy \\ \midrule
Latvia & Regional gov & Digital Centre & City of Ventspils \\ \midrule
Luxembourg & Govt & Luxembourg House of CyberSecurity OSPO & Ministry of the Economy \\ \midrule
Netherlands & Enterprise & Alliander OSPO & Alliander \\ \midrule
Netherlands & Govt & Bureau Open Source Software (BOSS) & Dutch Tax and Customs Administration \\ \midrule
Netherlands & Govt & OSPO for MinBZK & Ministry of the Interior \\ \midrule
Netherlands & Govt & Dutch Association of Municipalities and Regions (VNG) & Collaboration of Municipalities \\ \midrule
Netherlands & Regional gov & Province of Zeeland OSPO & Province of Zeeland \\ \midrule
Romania & NGO & Code for Romania* & *Not public sector \\ \midrule
Slovakia & Regional gov & Department for Digital Services and Innovation & City of Bratislava \\ \midrule
Sweden & Enterprise & SVT OSPO & Sveriges Television \\ \midrule
Sweden & Govt & DIGG (Digitisation Agency of Sweden) & DIGG (Digitisation Agency of Sweden) \\ \bottomrule
\end{tabular}
\end{table*}



\bibliographystyle{elsarticle-num} 
\bibliography{mybibliography}



\end{document}